\pdfoutput=1
\documentclass[11pt]{article}
\usepackage{subfig}
\usepackage{amssymb,amsmath,amsthm}
\usepackage{graphics,epsfig}
\usepackage{algorithmic}
\usepackage{algorithm}
\usepackage[margin=1 in]{geometry}

\usepackage{lipsum}
\usepackage{mathtools}
\usepackage{cuted}
\usepackage{hhline}
\usepackage{framed}
\usepackage{tikz}
\usepackage{url}

\newtheorem{defs}{Definition}
\newtheorem{lem}{Lemma}[section]

\newtheorem{coro}[lem]{Corollary}
\newtheorem{thm}[lem]{Theorem}
\newtheorem{obs}[lem]{\textbf{Observation}}

\newtheorem{prop}[lem]{Proposition}

\newtheorem{conj}{Conjecture}
\theoremstyle{definition}
\newtheorem{rem}[lem]{Remark}
\newtheorem{example}[lem]{Example}

\newcommand{\len}{\mathit{len}}
\newcommand{\spn}{\mathit{span}}
\newcommand{\SPAN}{\mathit{SPAN}}
\newcommand{\sav}{\mathit{sav}}
\newcommand{\seg}{\mathit{seg}}
\newcommand{\OPT}{\mathrm{OPT}}
\newcommand{\J}{\mathcal{J}}
\newcommand{\T}{\mathcal{T}}
\newcommand{\cDEG}{c_{\mathrm{DEG}}}
\newcommand{\cFLR}{c_{\mathrm{FLR}}}
\newcommand{\cMATCH}{c_{\mathrm{MATCH}}}
\newcommand{\cCFG}{c_{\mathrm{CFG}}}
\newcommand{\cCAT}{c_{\mathrm{CAT}}}
\newcommand{\cLOAD}{c_{\mathrm{LOAD}}}

\title{Minimizing Total Busy Time with Bounded Parallelism: A $3/2$-Approximation for Parallelism Two and a Forest Characterization for General Parallelism}
\author{Chi-Yeh~Chen 
\\ Department of Computer Science and Information
Engineering, \\ National Cheng Kung University, \\
Taiwan, ROC. \\
chency@mail.csie.ncku.edu.tw.}

\begin{document}

\maketitle
\begin{abstract}
We study the minimum total busy time problem: jobs with fixed processing intervals must be assigned to identical machines, each of which can process at most $g$ jobs at a time, so that the total time during which the machines are busy is minimized. The problem is NP-hard already for $g=2$, and, to the best of our knowledge, no approximation ratio below $2$ was known for general instances with $g=2$, where the ratio $2$ is attained by giving every job its own machine. We present a combinatorial $3/2$-approximation algorithm for $g=2$. The algorithm applies the two-phase local-ratio technique to the known pair formulation of the problem, processing the overlapping job pairs in order of the right ends of their overlaps. The cost of its schedule is at most $L-\lambda/2$, where $L$ is the total processing time and $\lambda$ is the optimal value of the linear relaxation of the pair formulation; hence the integrality gaps of this relaxation and of the configuration LP are at most $3/2$. The result extends to temporal bin packing with items larger than one third of the bin capacity. For general $g$, we show that optimal schedules correspond exactly to maximum-weight forests of a special kind, which we call $g$-caterpillar forests; this yields an exact integer program and a linear relaxation of polynomial size for every fixed $g$. We also observe that securing a $(1-1/g)$ fraction of the optimal savings suffices for a $(2-1/g)$-approximation. On random instances with $g=2$, evaluated with exact optima or tight lower bounds obtained from the pair formulation, the average ratio of the algorithm to the optimum is at most $1.075$ for $100$ jobs and at most $1.070$ for $200$ jobs.

\begin{keywords}
Approximation algorithms, Interval scheduling, Busy time, Local ratio, Linear programming, Temporal bin packing
\end{keywords}
\end{abstract}

\section{Introduction}\label{sec:introduction}
Interval scheduling is one of the basic models of scheduling theory (see, e.g.,~\cite{lawler93,Kolen07,Kovalyov07} and the references therein). Each job is tied to a fixed interval of time and must be processed throughout that interval, and in the classical model a machine processes at most one job at a time. This paper studies interval scheduling with \textit{bounded parallelism}, in which a single machine may process up to $g\geq 1$ jobs simultaneously. We are given a set $\J=\{J_1,\ldots,J_n\}$ of $n$ jobs. Job $J_j$ has a start time $s_j$ and a completion time $c_j>s_j$, and it must be processed during the interval $I_j=[s_j,c_j)$. Every job is assigned to one machine from an unbounded supply of identical machines, and an assignment is feasible if at no point in time a machine processes more than $g$ jobs. A machine is \textit{busy} at time $t$ if at least one of its jobs is processed at time $t$; otherwise it is \textit{idle} at $t$. The \textit{minimum total busy time problem} asks for a feasible schedule that minimizes the total busy time of all machines~\cite{Flammini10,Mertzios15}. Without loss of generality the busy time of every machine is one contiguous interval, because a machine whose busy time consists of several separated periods can be split into several machines without changing the total busy time.

\begin{figure}[t]
    \centering
\begin{tikzpicture}[x=1.1cm,y=0.5cm]
\draw[->,gray] (-0.2,0) -- (7.3,0) node[right]{\small time};
\foreach \s/\e/\y/\n in {0/1.5/7/1, 0.5/3.5/6/2, 0.3/2.5/5/3, 1.8/3.0/4/4, 3.1/4.2/3/5, 4.8/6.2/2/6, 5.2/6.8/1/7}{
  \draw[very thick] (\s,\y) -- (\e,\y);
  \node[left] at (\s,\y) {\small $J_{\n}$};
}
\begin{scope}[xshift=9.2cm,yshift=0cm,every node/.style={circle,draw,fill=gray!15,inner sep=1.5pt,minimum size=17pt}]
\node (1) at (1,7) {\small $J_1$};
\node (2) at (0.2,5.2) {\small $J_2$};
\node (3) at (1.8,5.2) {\small $J_3$};
\node (5) at (0.2,3.2) {\small $J_5$};
\node (4) at (1.8,3.2) {\small $J_4$};
\node (6) at (0.2,1.0) {\small $J_6$};
\node (7) at (1.8,1.0) {\small $J_7$};
\draw (1)--(2) (1)--(3) (2)--(3) (2)--(5) (2)--(4) (3)--(4) (6)--(7);
\end{scope}
\end{tikzpicture}
    \caption{Seven jobs $J_1,\ldots,J_7$, drawn as their processing intervals (left), and the corresponding interval graph (right).}
    \label{fig:interval}
\end{figure}
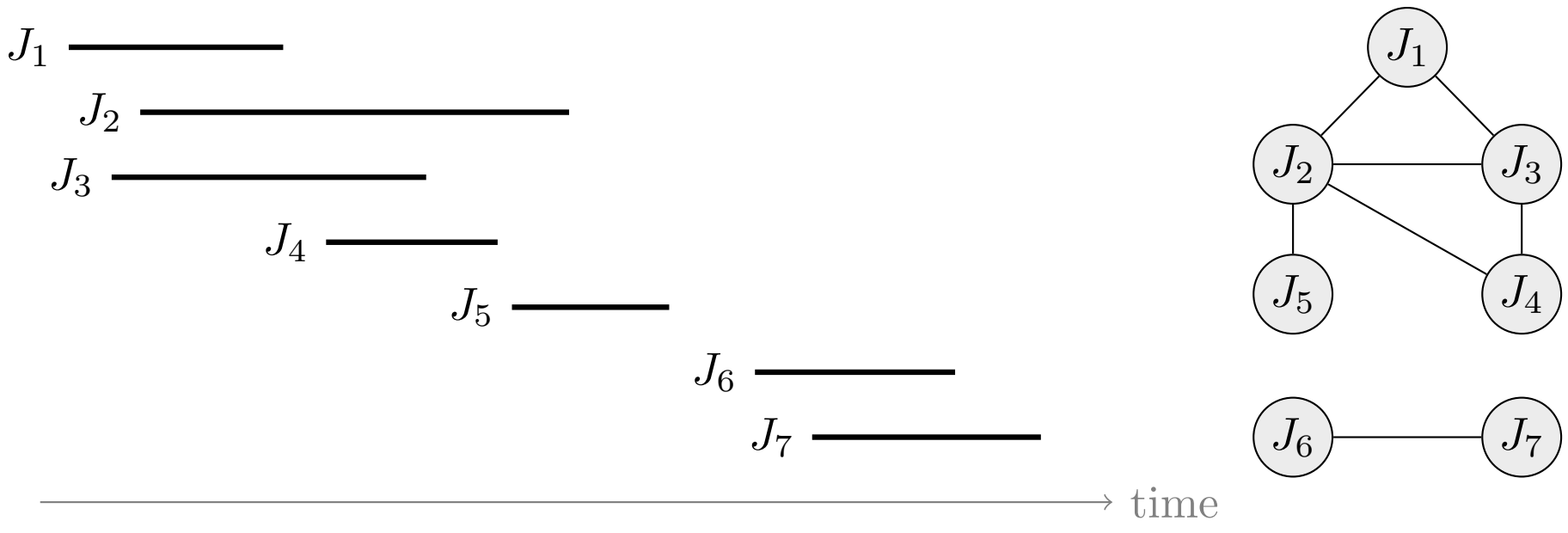

An instance can be represented by an \textit{interval graph}, the intersection graph of the job intervals, which has one vertex per job and an edge between two jobs whenever their intervals overlap~\cite{Flammini10}. Figure~\ref{fig:interval} shows an example. Jobs in different connected components of the interval graph are never processed at the same time, so the components can be scheduled independently; hence we may assume that the interval graph is connected. Two special classes of instances have received particular attention. An instance is \textit{proper} if no job interval properly contains another one, and it is a \textit{clique instance} if all job intervals share a common point, that is, if its interval graph is complete.

The case $g=2$ is the smallest nontrivial case. It is already NP-hard~\cite{Winkler03}, yet, as far as we know, the best approximation ratio known for it on general instances is $2$, the same as for every other $g$~\cite{ChangKM17}. For $g=2$ this ratio is attained by the trivial schedule that places every job on its own machine, because every machine processes at most two jobs at a time and therefore any schedule costs at least half of the total processing time. In this paper we look at the problem through \textit{savings}: the busy time of a machine equals the total length of its jobs minus the time saved by processing jobs in parallel, so minimizing the busy time is the same as maximizing the total savings. For $g=2$ the savings of a machine are exactly the total overlap of the pairs of jobs that it processes together, and the problem becomes a packing problem over overlapping pairs; this pair formulation is due to Muir \textit{et al.}~\cite{Muir24} (see also~\cite{MartinovicS25}). We show that the pair formulation has enough structure for the local-ratio technique~\cite{BarNoy01} to give a $3/2$-approximation, and we extend the structural description to every $g$.

\subsection{Applications}
\textbf{Optical network design.} In optical networks that use wavelength-division multiplexing (WDM), communication between two nodes is carried by a lightpath, and lightpaths that share a fiber link must use different wavelengths~\cite{Flammini08,Flammini11}. In the fiber minimization problem of Winkler and Zhang~\cite{Winkler03}, the network is a path, every lightpath is an interval, a fiber carries at most $g$ wavelengths, and the cost of a fiber is its length; this is exactly the busy time problem with fibers as machines. Related cost models charge for optical add-drop multiplexers (OADMs) or for regenerators, where traffic grooming allows up to $g$ lightpaths to share one device~\cite{Flammini08,Saradhi10,Flammini11,FlamminiAlgo14}; on path topologies these models again reduce to busy time minimization with the grooming factor as the machine capacity~\cite{Flammini10,Khandekar15}.

\textbf{Power-aware scheduling and cloud computing.} The energy consumed by a server is, to a first approximation, proportional to the time during which it is switched on. If each server can host at most $g$ jobs at a time, minimizing the energy of a cluster amounts to minimizing the total busy time~\cite{Vasic09,carli14,Khandekar15}. The same model describes the placement of virtual machines with known start and end times on physical hosts~\cite{TianYeo15,ChangKM17}. When jobs have heterogeneous resource demands, it becomes the MinUsageTime dynamic bin packing problem~\cite{Li14,LiuTang22b} and the busy time variant of temporal bin packing studied in operations research~\cite{deCauwer16,DellAmico20,Aydin20,Martinovic23,MartinovicS25}. In particular, temporal bin packing with half-capacity jobs (TBP2), studied by Muir, Marshall and Toriello~\cite{Muir24}, has items whose sizes lie in $(1/3,1/2]$ of the bin capacity, so every bin holds at most two items at a time; it is the case $g=2$ of our problem.

\subsection{Related Work}
\textbf{Interval jobs.} Winkler and Zhang~\cite{Winkler03} introduced the problem in the context of fiber minimization and proved that it is NP-hard for $g=2$. Flammini \textit{et al.}~\cite{Flammini10} showed that the greedy algorithm FirstFit, which considers the jobs in order of non-increasing length, is a $4$-approximation; they also gave $2$-approximations for proper instances and for clique instances, and a $(2+\epsilon)$-approximation for instances with bounded job lengths. Alicherry and Bhatia~\cite{AlicherryBhatia03} and, independently, Kumar and Rudra~\cite{KumarRudra05} designed algorithms for closely related problems in optical line system design whose cost is at most twice the load bound $\int\lceil N(t)/g\rceil\,dt$, where $N(t)$ is the number of jobs at time $t$; Chang, Khuller and Mukherjee~\cite{ChangKM17} observed that these results give $2$-approximations for the busy time problem, and they also presented the combinatorial algorithm GreedyTracking, a $3$-approximation that extends to jobs with flexible windows. Andrews and Zhang~\cite{AndrewsZhang05} studied fiber minimization in general network topologies, where fibers are counted separately on every link; this objective differs from busy time. Mertzios \textit{et al.}~\cite{Mertzios15} gave a $\frac{g\cdot H_g}{H_g+g-1}$-approximation for clique instances, where $H_g$ is the $g$-th harmonic number, and a $(2-1/g)$-approximation for proper instances; they also studied the maximization of the number of scheduled jobs under a busy time budget. For clique instances with $g=2$, every machine holds at most two jobs, and Mertzios \textit{et al.} solved this case exactly by a maximum-weight matching in which the weight of a pair of jobs is the length of their overlap. The pair formulation extends this idea to general instances, in which a job may share its machine with different jobs at different times. Khandekar \textit{et al.}~\cite{khandekar10,Khandekar15} considered jobs with capacity demands and release-time/deadline windows, gave a $5$-approximation, and obtained a polynomial-time approximation scheme for clique instances and an exact algorithm for laminar instances of unit demands. Koehler and Khuller~\cite{KoehlerKhuller17} studied the problem with a bounded number of machines. To the best of our knowledge, for general instances the ratio $2$ has not been improved for any fixed $g\geq 2$.

\textbf{Exact methods and temporal bin packing.} The operations research literature treats the problem as temporal bin packing with the busy time objective. Muir \textit{et al.}~\cite{Muir24} studied TBP2 and proposed an integer programming formulation that views the problem as a sequence of linked matching problems; they compared a hierarchy of lower bounds, including a clique bound and several linear relaxations, proved that two of these relaxations have the same value (Section~\ref{sec:lps}), and showed that the optimal schedule costs at most $(2-1/m)$ times the clique bound, where $m$ is the number of maximal cliques. Martinovic and Strasdat~\cite{MartinovicS25,MartinovicS25b} developed assignment, flow and savings-based integer programs and heuristics for busy time minimization with general capacities. Branch-and-price, flow-based and multi-objective approaches for temporal bin packing with other objectives can be found in~\cite{deCauwer16,DellAmico20,Aydin20,Martinovic23}. These works are computational; to the best of our knowledge, they do not give approximation guarantees below $2$ for the busy time objective.

\textbf{Further variants.} Shalom \textit{et al.}~\cite{Shalom14} studied the online version with interval jobs. Online algorithms for flexible jobs were given by Albers and van der Heijden~\cite{AlbersH24}, and Călinescu \textit{et al.}~\cite{Calinescu24} studied online flexible busy time scheduling on heterogeneous machines. Ren and Tang~\cite{RenTang20} and Liu and Tang~\cite{LiuTang22} considered heterogeneous machine types with different capacities and cost rates, and Liu and Tang~\cite{LiuTang22b} studied dynamic bin packing with predictions. The closely related active time problem, in which a single machine with capacity $g$ pays for every time slot in which it is on, was studied by Chang \textit{et al.}~\cite{ChangKM17} and Cao \textit{et al.}~\cite{Cao22}, who gave a $9/5$-approximation for nested instances. Călinescu and Liu~\cite{CalinescuLiu26} recently considered busy time minimization with preemption and migration.

\textbf{Techniques.} Our algorithm uses the two-phase local-ratio technique of Bar-Noy \textit{et al.}~\cite{BarNoy01}, which they applied to resource allocation and interval scheduling. Its analysis follows the inductive-independence argument of Ye and Borodin~\cite{YeBorodin12}: if the elements can be ordered so that the later neighbours of every element can be covered by $k$ cliques, the method gives a $1/k$-approximation for the maximum-weight independent set. Surveys of related scheduling models can be found in~\cite{Leung04,Brucker07}.

\subsection{Our Contributions}
We take the pair formulation of the case $g=2$ (Lemma~\ref{lem:pair}) and the equality of the pair relaxation and the configuration LP from Muir \textit{et al.}~\cite{Muir24} as our starting point. Our contributions are the following.
\begin{itemize}
\item For $g=2$, we present LR-PAIR, a combinatorial algorithm that runs in $O(n^2\log n)$ time and returns a schedule of cost at most $L-\lambda/2\le\frac{3}{2}\,\OPT$, where $L$ is the total processing time and $\lambda$ is the optimal value of the linear relaxation of the pair formulation (Theorem~\ref{thm:main}). To the best of our knowledge, this is the first approximation ratio below $2$ for general instances with $g=2$. The guarantee is instance-sensitive: the ratio is at most $(1-\rho/2)/(1-\rho)$ with $\rho=\lambda/L$.
\item As corollaries, the integrality gap of the pair relaxation and of the configuration LP is at most $3/2$ for $g=2$; a simple example shows that it is at least $4/3$ (Corollary~\ref{cor:gap}). For this we give a short self-contained proof that the pair relaxation is not stronger than the configuration LP (Lemma~\ref{lem:cfg}). The algorithm and its guarantee extend to temporal bin packing with item sizes larger than one third of the capacity (Corollary~\ref{cor:tbp}).
\item For general $g$, we show that the optimal savings equal the maximum weight of a $g$-caterpillar forest (Theorem~\ref{thm:cat}). This gives an exact integer program (Proposition~\ref{prop:ip}) and a linear relaxation of polynomial size for every fixed $g$, which does not need column generation; for $g=2$ the relaxation lies between the floor-cut and the matching relaxations (Proposition~\ref{prop:g2}).
\item We show that an algorithm that secures a $\beta$ fraction of the optimal savings is a $(1+(1-\beta)(g-1))$-approximation (Proposition~\ref{prop:beta}). Hence a savings guarantee of $1-1/g$ would give the ratio $2-1/g$; Theorem~\ref{thm:main} is the case $g=2$. We conjecture that a $(2-1/g)$-approximation exists for every $g$.
\item We evaluate the algorithm on random instances. Using the pair formulation, we solve all instances with $100$ jobs and most instances with $200$ jobs to optimality. LR-PAIR is on average within $7.5\%$ ($n=100$) and at most $7.0\%$ ($n=200$) of the optimum, compared with $8.4\%$ and $8.5\%$ for GreedyTracking, and its largest ratio is $1.17$. We also study the effect of tie-breaking and of instances in general position on the worst case of LR-PAIR.
\item We point out a gap in the proof of the clique-instance bound of Mertzios \textit{et al.}~\cite{Mertzios15} (Appendix~\ref{app:mertzios}).
\end{itemize}

\subsection{Organization}
Section~\ref{sec:Preliminaries} introduces notation and the savings view of the problem. Section~\ref{sec:pairs} presents the pair formulation for $g=2$ and its linear relaxations. Section~\ref{sec:algorithm} describes LR-PAIR and proves its approximation ratio. Section~\ref{sec:general} treats general $g$. Section~\ref{sec:Results} reports experimental results, and Section~\ref{sec:Conclusion} concludes. Appendix~\ref{app:mertzios} discusses the clique-instance bound of Mertzios \textit{et al.}, Appendix~\ref{app:gen} lists the instance generators, and Appendix~\ref{app:inst} lists the bad instances found by local search.

\section{Notation and Preliminaries}\label{sec:Preliminaries}
A set of jobs $\J$ is scheduled on a set $\mathcal{M}$ of identical machines with \textit{parallelism parameter} $g$, the maximum number of jobs that a machine can process at the same time. Jobs and their intervals are used interchangeably. We use half-open intervals, so a job that completes at time $t$ and a job that starts at time $t$ are never processed at the same time; two intervals \textit{overlap} if their intersection has positive length.

\begin{defs}
The length of an interval $I=[s,c)$ is $\len(I)=c-s$. For a set $\mathcal{I}$ of intervals, $\len(\mathcal{I})=\sum_{I\in\mathcal{I}}\len(I)$.
\end{defs}

\begin{defs}
For a set $\mathcal{I}$ of intervals, $\SPAN(\mathcal{I})=\bigcup_{I\in\mathcal{I}}I$ and $\spn(\mathcal{I})=\len(\SPAN(\mathcal{I}))$, the total length of the union.
\end{defs}

Clearly $\spn(\mathcal{I})\le\len(\mathcal{I})$, with equality if and only if the intervals of $\mathcal{I}$ are pairwise non-overlapping. A \textit{schedule} is a function $s:\J\to\mathcal{M}$. Let $\J_i=s^{-1}(M_i)$ be the set of jobs assigned to machine $M_i$. The schedule is \textit{valid} if for every machine $M_i$ and every time $t$, at most $g$ jobs of $\J_i$ contain $t$. The cost of machine $M_i$ is its busy time $\mathit{busy}_i=\spn(\J_i)$, and the cost of the schedule is $\mathit{cost}^s=\sum_i \mathit{busy}_i$. An optimal schedule is denoted by $s^*$ and its cost by $\mathit{cost}^*$, which we also write as $\OPT$; when $g$ needs to be explicit we write $\mathit{cost}^*_g$. For a time $t$, let $\J(t)=\{J_j\in\J: t\in I_j\}$ be the set of jobs \textit{running} at $t$ and $N(t)=|\J(t)|$; for a subset $\mathcal{A}\subseteq\J$, $N_{\mathcal{A}}(t)$ denotes the number of jobs of $\mathcal{A}$ running at $t$. Finally, $L=\len(\J)$ denotes the total processing time.

\textbf{The order $\prec$.} We write $J_a\prec J_b$ if $s_a<s_b$, or $s_a=s_b$ and $c_a>c_b$, or $s_a=s_b$, $c_a=c_b$ and $a<b$; that is, jobs are ordered by start time, then by decreasing completion time, then by index. Since ties are broken by index, $\prec$ is a strict total order on $\J$; for a nonempty set $S\subseteq\J$, the \textit{$\prec$-smallest} (resp.\ \textit{$\prec$-largest}) job of $S$ is its minimum (resp.\ maximum) element in this order. If $J_a\prec J_b$ and the two jobs overlap, then $J_a$ is running at time $s_b$.

The following bounds are standard.

\begin{obs}[\cite{Flammini10,Mertzios15}]\label{obs:bounds}
For any instance $\J$, parallelism parameter $g\ge1$ and valid schedule $s$:
\begin{itemize}
\item the parallelism bound: $\mathit{cost}^s\ge \len(\J)/g$;
\item the span bound: $\mathit{cost}^s\ge\spn(\J)$;
\item the length bound: $\mathit{cost}^s\le \len(\J)$.
\end{itemize}
\end{obs}

The problem is defined as follows.
\begin{framed}
\noindent\textbf{The minimum total busy time problem}\\
\textbf{Input:} $(\J,g)$, where $\J$ is a set of jobs with intervals $I_j=[s_j,c_j)$ and $g$ is the parallelism parameter.\\
\textbf{Output:} A valid schedule $s:\J\to\mathcal{M}$.\\
\textbf{Objective:} Minimize $\mathit{cost}^s$.
\end{framed}

\textbf{Elementary segments.} Let $p_1<p_2<\cdots<p_K$ be the distinct values in $\{s_j,c_j: J_j\in\J\}$. The intervals $[p_k,p_{k+1})$, $1\le k<K$, are the \textit{elementary segments}, and $\T$ denotes their set; $|\T|<2n$. The set of running jobs is constant on every $T\in\T$, and we write $\J(T)$, $N(T)$ and $|T|$ for this set, its size and the length of $T$. Every job interval and every intersection of job intervals is a union of elementary segments.

\textbf{Savings.} The \textit{savings} of a set $\mathcal{A}\subseteq\J$ of jobs are
\begin{equation}\label{eq:sav}
\sav(\mathcal{A})=\len(\mathcal{A})-\spn(\mathcal{A})=\int\bigl(N_{\mathcal{A}}(t)-1\bigr)^+\,dt ,
\end{equation}
where $x^+=\max(x,0)$ denotes the positive part of $x$ and $[\cdot]$ denotes the indicator of a condition (it equals $1$ if the condition holds and $0$ otherwise). The second equality holds because $(N-1)^+=N-[N\ge1]$ for every integer $N\ge0$, $\int N_{\mathcal{A}}(t)\,dt=\len(\mathcal{A})$ and $\int[N_{\mathcal{A}}(t)\ge1]\,dt=\spn(\mathcal{A})$. Consequently, for every schedule $s$,
\begin{equation}\label{eq:costsav}
\mathit{cost}^s=L-\sum_{i}\sav(\J_i),
\end{equation}
and minimizing the total busy time is equivalent to maximizing the total savings. Let $S^*=L-\OPT$ denote the optimal savings. Table~\ref{tab:notations} summarizes the notation.

\begin{table}[ht]
\caption{Notation and Terminology}
    \centering
        \begin{tabular}{||c|p{3.75in}||}
    \hline
$\J$, $n$ & The set of jobs and its size. \\ \hline
$I_j=[s_j,c_j)$ & The interval of job $J_j$. \\ \hline
$g$ & The parallelism parameter. \\ \hline
$\len(\cdot)$, $\spn(\cdot)$ & Total length and length of the union of a set of intervals. \\ \hline
$L$ & The total processing time $\len(\J)$. \\ \hline
$\J(t)$, $N(t)$ & The set and the number of jobs running at time $t$. \\ \hline
$\T$ & The set of elementary segments. \\ \hline
$\mathit{cost}^s$, $\mathit{cost}^*=\OPT$ & The cost of schedule $s$ and the optimal cost. \\ \hline
$\sav(\mathcal{A})$, $S^*$ & The savings of a job set and the optimal total savings $L-\OPT$. \\ \hline
$\prec$ & Order by start time, then decreasing completion time, then index. \\ \hline
$e=\{J_a,J_b\}$, $\seg(e)$, $w(e)$ & An overlap edge, its segment $I_a\cap I_b=[\sigma_e,\tau_e)$ and its weight $\tau_e-\sigma_e$ ($g=2$). \\ \hline
$E$, $m$ & The set of overlap edges and its size. \\ \hline
$\lambda$, $\cDEG$ & The optimal value of the pair relaxation (P) and the lower bound $L-\lambda$. \\ \hline
$\cFLR$, $\cMATCH$, $\cCFG$, $\cCFG^+$, $\cCAT$ & Lower bounds from further relaxations (Sections~\ref{sec:lps} and~\ref{sec:compact}). \\ \hline
        \end{tabular}
    \label{tab:notations}
\end{table}

\section{The Pair Formulation for $g=2$}\label{sec:pairs}
Throughout this section and the next, $g=2$.

\begin{defs}
If the intervals of two jobs $J_a$ and $J_b$ overlap, the pair $e=\{J_a,J_b\}$ is an \textit{overlap edge}. Its \textit{segment} is $\seg(e)=I_a\cap I_b=[\sigma_e,\tau_e)$ and its \textit{weight} is $w(e)=\tau_e-\sigma_e$. Let $E$ be the set of overlap edges and $m=|E|\le n(n-1)/2$. Two distinct edges \textit{conflict} if they share a job and their segments intersect. A set $F\subseteq E$ is \textit{compatible} if no two edges of $F$ conflict, and $w(F)=\sum_{e\in F}w(e)$.
\end{defs}

Two edges $\{J_a,J_b\}$ and $\{J_a,J_c\}$ conflict exactly when the three jobs $J_a,J_b,J_c$ are running at a common time. The following lemma, which is due to Muir \textit{et al.}~\cite[Prop.~5.1]{Muir24} in an equivalent form based on maximal cliques, shows that compatible edge sets and valid schedules are two descriptions of the same objects. The savings model of Martinovic and Strasdat~\cite{MartinovicS25} for $g=2$ rests on the same observation, and for clique instances, where compatible sets are exactly the matchings, it reduces to the matching algorithm of Mertzios \textit{et al.}~\cite[Lemma~3.1]{Mertzios15}. We include a proof because the parent structure used in part~(b) is generalized in Section~\ref{sec:general}.

\begin{lem}[Pair formulation~\cite{Muir24}]\label{lem:pair}
Let $g=2$.
\begin{itemize}
\item[(a)] For every valid schedule $s$, the set $F(s)$ of overlap edges whose two jobs are assigned to the same machine is compatible, and $\mathit{cost}^s=L-w(F(s))$.
\item[(b)] For every compatible set $F$, the schedule $s_F$ that uses one machine for each connected component of the graph $(\J,F)$ is valid, and $\mathit{cost}^{s_F}=L-w(F)$.
\end{itemize}
Consequently, $\OPT=L-\max\{w(F): F\subseteq E\text{ compatible}\}$.
\end{lem}

\begin{proof}
(a) If two edges $\{J_a,J_b\},\{J_a,J_c\}\in F(s)$ had segments that intersect at time $t$, the three jobs $J_a,J_b,J_c$ would run on the same machine at time $t$, contradicting validity. Since each machine $M_i$ runs at most two jobs at a time, \eqref{eq:sav} gives $\sav(\J_i)=\int[N_{\J_i}(t)=2]\,dt$, and every time $t$ with $N_{\J_i}(t)=2$ belongs to the segment of exactly one edge of $F(s)$ inside $\J_i$. Hence $\sav(\J_i)=\sum_{e\in F(s),\,e\subseteq\J_i}w(e)$, and \eqref{eq:costsav} gives the claim.

(b) For an edge $\{J_a,J_b\}\in F$ with $J_a\prec J_b$, call $J_a$ the \textit{parent} of $J_b$. Recall that if $J_a\prec J_b$ and the two jobs overlap, then $J_a$ is running at time $s_b$.

\emph{(b1) Every job has at most one parent.} If $J_{a_1}$ and $J_{a_2}$ were both parents of $J_b$, both segments would contain $s_b$, so the two edges would conflict. Hence $F$ is a forest, and the root of each tree is its $\prec$-smallest job.

\emph{(b2) In every connected component $K$ of $(\J,F)$, at most two jobs are running at any time.} Suppose not. Call a job $J_y\in K$ \emph{crowded} if at time $s_y$ at least two jobs of $K$ that precede $J_y$ are running. If three jobs of $K$ run at time $t$, the $\prec$-largest of them is crowded, so crowded jobs exist; let $J_y$ be the $\prec$-smallest one. Exactly two jobs of $K$ that precede $J_y$ run at time $s_y$, since if there were three, the $\prec$-largest of them would be a crowded job preceding $J_y$; call them $J_x$ and $J_z$, where $J_x\prec J_z$. Then $J_x$ runs at time $s_z$, and $J_z$ is not a root (since $J_x\in K$ precedes it); since $J_z$ is not crowded (as $J_z\prec J_y$), the only job of $K$ preceding $J_z$ that runs at $s_z$ is $J_x$, so $J_x$ is the parent of $J_z$ and $\{J_x,J_z\}\in F$. The job $J_y$ is not the root of its tree, so its parent runs at time $s_y$ and precedes $J_y$; hence the parent is $J_x$ or $J_z$. In either case, the edge joining $J_y$ to its parent and the edge $\{J_x,J_z\}$ share a job and their segments contain $s_y$, a contradiction.

\emph{(b3) Any two overlapping jobs $J_u\prec J_v$ of the same component are joined by an edge of $F$.} The job $J_v$ is not a root (since $J_u$ precedes it in the same tree), so its parent runs at time $s_v$; so does $J_u$, because it overlaps $J_v$ and precedes it. If the parent and $J_u$ were different, three jobs of the component would run at time $s_v$, contradicting (b2); hence $J_u$ is the parent of $J_v$.

By (b2) the schedule $s_F$ is valid. By (b3) and the computation in (a), the savings of each component equal the total weight of its $F$-edges, and \eqref{eq:costsav} gives $\mathit{cost}^{s_F}=L-w(F)$.
\end{proof}

\subsection{Linear relaxations}\label{sec:lps}
For an edge $e$ and an elementary segment $T$, $T\subseteq\seg(e)$ holds if and only if the segment of $e$ intersects $T$. The \textit{pair relaxation} is
\begin{equation*}
(\mathrm{P})\qquad \lambda=\max\ \sum_{e\in E}w(e)\,y_e\quad\text{s.t.}\quad \sum_{e\ni J_a,\ T\subseteq\seg(e)}y_e\le1\quad(\forall J_a\in\J,\ \forall T\in\T),\qquad y\ge0 .
\end{equation*}
It has $m$ variables and fewer than $2n^2$ constraints. We write $\cDEG=L-\lambda$, following the notation of Muir \textit{et al.}~\cite{Muir24}.

\begin{lem}\label{lem:lp}
$\cDEG\le\OPT$ and $\lambda\le L/2$.
\end{lem}
\begin{proof}
The incidence vector of an optimal compatible set is feasible for (P), because two edges that contain $J_a$ and whose segments contain $T$ conflict; by Lemma~\ref{lem:pair}, $\lambda\ge S^*$. For the second claim, fix $T\in\T$ and add the constraints of all jobs in $\J(T)$. Every edge $e$ with $T\subseteq\seg(e)$ appears twice, so $2\sum_{e:T\subseteq\seg(e)}y_e\le N(T)$. Therefore $\lambda=\sum_{T\in\T}|T|\sum_{e:T\subseteq\seg(e)}y_e\le\sum_{T}|T|\,N(T)/2=L/2$.
\end{proof}

Two families of valid inequalities strengthen (P). The \textit{floor cuts} $\sum_{e:T\subseteq\seg(e)}y_e\le\lfloor N(T)/2\rfloor$ express that at most $\lfloor N(T)/2\rfloor$ disjoint pairs can run during $T$; we write $\cFLR$ for the resulting bound. The \textit{blossom inequalities} $\sum_{e\subseteq S,\,T\subseteq\seg(e)}y_e\le(|S|-1)/2$ for all $T\in\T$ and all odd sets $S\subseteq\J(T)$ describe, together with (P), the matching polytope of the jobs running during $T$~\cite{Edmonds65}; they can be separated in polynomial time~\cite{PadbergRao82}, and with them we obtain the bound $\cMATCH$ studied in~\cite{Muir24}. Since the floor cut for odd $N(T)$ is the blossom inequality with $S=\J(T)$, and for even $N(T)$ it is implied by (P), we have $\cDEG\le\cFLR\le\cMATCH\le\OPT$.

A different relaxation is the \textit{configuration LP}. A \textit{configuration} is a nonempty set $C\subseteq\J$ of jobs that can be processed by one machine, that is, at most $g$ jobs of $C$ run at any time; its cost is $\spn(C)$. Configurations need not be connected, but a disconnected configuration costs exactly as much as its connected parts. The configuration LP is
\begin{equation*}
\cCFG=\min\ \sum_{C}\spn(C)\,x_C\quad\text{s.t.}\quad\sum_{C\ni J_j}x_C=1\quad(\forall J_j\in\J),\qquad x\ge0 .
\end{equation*}
It has exponentially many variables and is solved by column generation in practice~\cite{Muir24,MartinovicS25}; Muir \textit{et al.}\ call its value $c_{\mathrm{PART}}$. It can be strengthened by the \textit{time cuts} $\sum_{C:\,C\cap\J(T)\neq\emptyset}x_C\ge\lceil N(T)/g\rceil$ for all $T\in\T$, which state that at least $\lceil N(T)/g\rceil$ machines are busy during $T$; we denote the strengthened value by $\cCFG^+$. Finally, the \textit{load bound} is $\cLOAD=\sum_{T\in\T}|T|\lceil N(T)/g\rceil$; it is implied by the time cuts, so $\cLOAD\le\cCFG^+$.

Muir \textit{et al.}~\cite{Muir24} proved that $\cCFG=\cDEG$ for $g=2$. We only need one direction, which has a short proof.

\begin{lem}\label{lem:cfg}
For $g=2$, $\cDEG\le\cCFG\le\cCFG^+\le\OPT$.
\end{lem}
\begin{proof}
Only the first inequality needs proof. Let $x$ be feasible for the configuration LP, and for every edge $e$ let $y_e=\sum_{C\supseteq e}x_C$. Lemma~\ref{lem:pair}(a), applied to a schedule that contains the machine $C$, shows $\sav(C)=\sum_{e\subseteq C}w(e)$. Hence
\[
\sum_C\spn(C)\,x_C=\sum_C\bigl(\len(C)-\sav(C)\bigr)x_C=\sum_{J_j}\len(I_j)\sum_{C\ni J_j}x_C-\sum_e w(e)\sum_{C\supseteq e}x_C=L-\sum_e w(e)\,y_e .
\]
The vector $y$ is feasible for (P): for a job $J_a$ and a segment $T$, every configuration $C\ni J_a$ contains at most one edge $e\ni J_a$ with $T\subseteq\seg(e)$, because at most two jobs of $C$ run during $T$; hence $\sum_{e\ni J_a,\,T\subseteq\seg(e)}y_e\le\sum_{C\ni J_a}x_C=1$. Therefore $\sum_C\spn(C)\,x_C\ge L-\lambda=\cDEG$.
\end{proof}

\section{A $3/2$-Approximation Algorithm for $g=2$}\label{sec:algorithm}
Our algorithm, LR-PAIR (Algorithm~\ref{alg:lr}), applies the two-phase local-ratio method~\cite{BarNoy01} to the edges of $E$. The edges are processed in order of non-decreasing right end $\tau_e$ of their segments. For the $k$-th edge $e_k$ in this order, let
\[
N^+(e_k)=\{e_k\}\cup\{e_l: l>k \text{ and } e_l \text{ conflicts with } e_k\}.
\]
In the first phase, each edge with positive residual weight is pushed on a stack, and its residual weight is subtracted from the residual weights of all edges in $N^+(e_k)$. In the second phase the stack is unwound, and an edge is kept if it does not conflict with an edge kept before. The algorithm does not solve a linear program; the relaxation (P) appears only in the analysis.

\begin{algorithm}[htbp]
\caption{LR-PAIR$(\J)$}
\begin{algorithmic}[1]
\REQUIRE A set $\J$ of jobs ($g=2$).
\ENSURE A valid schedule.
\STATE Compute the set $E$ of overlap edges and sort it as $e_1,\ldots,e_m$ such that $\tau_{e_1}\le\tau_{e_2}\le\cdots\le\tau_{e_m}$ (ties broken by a fixed rule, see Section~\ref{sec:tight}).
\STATE $r(e)\leftarrow w(e)$ for every $e\in E$; $\Sigma\leftarrow$ an empty stack.
\FOR{$k=1,2,\ldots,m$}
  \IF{$r(e_k)>0$}
    \STATE $\delta_k\leftarrow r(e_k)$; push $e_k$ onto $\Sigma$.
    \STATE $r(f)\leftarrow r(f)-\delta_k$ for every $f\in N^+(e_k)$.
  \ENDIF
\ENDFOR
\STATE $F\leftarrow\emptyset$.
\WHILE{$\Sigma$ is not empty}
  \STATE Pop an edge $e$ from $\Sigma$; if $e$ conflicts with no edge of $F$, then $F\leftarrow F\cup\{e\}$.
\ENDWHILE
\RETURN the schedule $s_F$ that uses one machine per connected component of $(\J,F)$.
\end{algorithmic}
\label{alg:lr}
\end{algorithm}

The key property of the order is that the later neighbours of an edge form two cliques of (P).

\begin{lem}[Two cliques]\label{lem:two}
Let $e_k=\{J_a,J_b\}$ with $\seg(e_k)=[\sigma,\tau)$, and let $T$ be the elementary segment that ends at $\tau$. Then $T\subseteq\seg(f)$ for every $f\in N^+(e_k)$. Consequently, $\sum_{f\in N^+(e_k)}y_f\le2$ for every feasible solution $y$ of (P).
\end{lem}
\begin{proof}
Let $f\in N^+(e_k)$. Since the segments of $f$ and $e_k$ intersect, $\sigma_f<\tau$; since $f$ does not precede $e_k$ in the order, $\tau_f\ge\tau$. Hence $\seg(f)\supseteq[\max(\sigma,\sigma_f),\tau)$. Both $\sigma$ and $\sigma_f$ are start times of jobs and are smaller than $\tau$, so they are at most the left end of $T$, and $T\subseteq\seg(f)$. Every edge of $N^+(e_k)$ contains $J_a$ or $J_b$. The edges that contain $J_a$ appear together in the constraint of $(J_a,T)$ in (P), so their $y$-values sum to at most $1$; the same holds for $J_b$.
\end{proof}

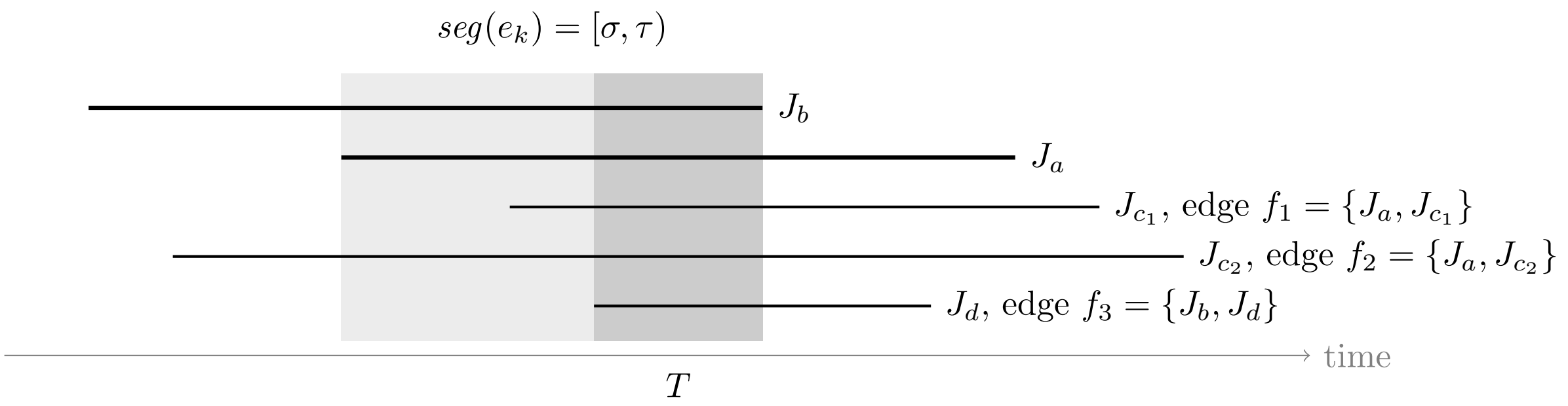
\begin{figure}[t]
\centering
\begin{tikzpicture}[x=0.85cm,y=0.5cm]
\draw[->,gray] (0,0) -- (15.5,0) node[right]{\small time};
\fill[gray!15] (4,0.3) rectangle (9,5.7);
\fill[gray!40] (7,0.3) rectangle (9,5.7);
\node at (8,-0.6) {\small $T$};
\draw[very thick] (1,5) -- (9,5) node[right]{\small $J_b$};
\draw[very thick] (4,4) -- (12,4) node[right]{\small $J_a$};
\draw[thick] (6,3) -- (13,3) node[right]{\small $J_{c_1}$, edge $f_1=\{J_a,J_{c_1}\}$};
\draw[thick] (2,2) -- (14,2) node[right]{\small $J_{c_2}$, edge $f_2=\{J_a,J_{c_2}\}$};
\draw[thick] (7,1) -- (11,1) node[right]{\small $J_{d}$, edge $f_3=\{J_b,J_{d}\}$};
\node at (6.5,6.6) {\small $\seg(e_k)=[\sigma,\tau)$};
\end{tikzpicture}
\caption{The edge $e_k=\{J_a,J_b\}$ has segment $[\sigma,\tau)$ (the whole shaded region, light and dark grey) and the earliest segment end among the edges shown. Every later edge that conflicts with $e_k$ contains the last elementary segment $T$ before $\tau$ (dark grey), so these edges form one clique of (P) at $J_a$ and one at $J_b$.}
\label{fig:two}
\end{figure}

Figure~\ref{fig:two} illustrates the lemma.

\begin{rem}[Relation to inductive independence]\label{rem:yb}
Lemma~\ref{lem:two} implies that, in the order of segment ends, the later conflicting neighbours of every edge lie in two cliques of the conflict graph; that is, the conflict graph has inductive independence number at most $2$, and the result of Ye and Borodin~\cite{YeBorodin12} gives a $\frac12$-approximation for the maximum-weight compatible set. For the approximation ratio alone this would suffice: savings of at least $S^*/2$ give $\mathit{cost}^s\le\OPT+S^*/2\le\frac32\OPT$, because $S^*=L-\OPT\le\OPT$ by the parallelism bound $L\le2\,\OPT$. Theorem~\ref{thm:main} adds two ingredients. The cliques are cliques of the relaxation (P), which gives the comparison with $\lambda$ instead of $S^*$; and the bound $\lambda\le L/2$ of Lemma~\ref{lem:lp} turns the comparison with $\lambda$ into a guarantee with respect to $\cDEG$. This yields the integrality-gap bounds of Corollary~\ref{cor:gap} and the instance-wise guarantee of Remark~\ref{rem:rho}.
\end{rem}

\begin{thm}\label{thm:main}
For $g=2$, LR-PAIR can be implemented to run in $O(m\log n)\subseteq O(n^2\log n)$ time and $O(n+m)$ space, and it returns a valid schedule $s$ with
\[
\mathit{cost}^s\ \le\ L-\frac{\lambda}{2}\ \le\ \frac32\,\cDEG\ \le\ \frac32\,\OPT .
\]
\end{thm}
\begin{proof}
The set $F$ is compatible by construction, so by Lemma~\ref{lem:pair}(b) the schedule $s_F$ is valid and $\mathit{cost}^{s_F}=L-w(F)$. Let $P$ be the set of indices of pushed edges.

\emph{Step 1: $w(F)\ge\sum_{k\in P}\delta_k$.} Let $f=e_l$ with $l\in P$. When $f$ is processed, its residual weight equals $w(f)$ minus the sum of $\delta_k$ over all $k\in P$, $k<l$, with $f\in N^+(e_k)$, and $\delta_l$ equals this residual weight. Since $f\in N^+(f)$ and $f\notin N^+(e_k)$ for $k>l$, we get $w(f)=\sum_{k\in P:\,f\in N^+(e_k)}\delta_k$. Summing over $f\in F\subseteq\{e_l:l\in P\}$ gives
\[
w(F)=\sum_{k\in P}\delta_k\,\bigl|F\cap N^+(e_k)\bigr| .
\]
Every pushed edge $e_k$ is either added to $F$ in the second phase or rejected because it conflicts with an edge $f\in F$ popped earlier. An edge popped earlier was pushed later, so $f$ comes after $e_k$ in the order and $f\in N^+(e_k)$. Hence $|F\cap N^+(e_k)|\ge1$ for all $k\in P$.

\emph{Step 2: $\lambda\le2\sum_{k\in P}\delta_k$.} After the first phase every edge has non-positive residual weight, so $w(f)\le\sum_{k\in P:\,f\in N^+(e_k)}\delta_k$ for every $f\in E$. For an optimal solution $y$ of (P), Lemma~\ref{lem:two} gives
\[
\lambda=\sum_{f}w(f)\,y_f\le\sum_{k\in P}\delta_k\sum_{f\in N^+(e_k)}y_f\le2\sum_{k\in P}\delta_k .
\]

Steps~1 and~2 give $w(F)\ge\lambda/2$, hence $\mathit{cost}^{s_F}\le L-\lambda/2=\cDEG+\lambda/2$. By Lemma~\ref{lem:lp}, $\lambda\le L/2\le L-\lambda=\cDEG$, which yields $\mathit{cost}^{s_F}\le\frac32\cDEG\le\frac32\OPT$.

\emph{Running time and space.} The $m=O(n^2)$ edges can be generated in $O(n^2)$ time and sorted in $O(m\log n)$ time. A direct implementation of line~6 of Algorithm~\ref{alg:lr} updates $|N^+(e_k)|\le2n-3$ residual weights per pushed edge and takes $O(nm)$ time. Instead, we compute the residual weight of an edge only when it is processed. Let $f=\{J_a,J_b\}$ be processed after a pushed edge $e_k$. Since $\sigma_{e_k}<\tau_{e_k}\le\tau_f$, the segments of $e_k$ and $f$ intersect if and only if $\tau_{e_k}>\sigma_f$. Hence $f\in N^+(e_k)$ if and only if $e_k$ contains $J_a$ or $J_b$ and $\tau_{e_k}>\sigma_f$, and no pushed edge other than $f$ itself contains both jobs. Therefore the residual weight of $f$ when it is processed is
\[
w(f)-D_a(\sigma_f)-D_b(\sigma_f),
\]
where $D_a(x)$ is the sum of $\delta_k$ over the pushed edges $e_k$ that contain $J_a$ and satisfy $\tau_{e_k}>x$. For every job we keep a Fenwick tree indexed by the segment ends of the edges that contain the job; then $D_a(x)$ is a suffix sum, and both queries and updates take $O(\log n)$ time. The first phase thus takes $O(m\log n)$ time and produces the same pushed edges and values $\delta_k$ as Algorithm~\ref{alg:lr}. In the second phase, the chosen edges that contain a fixed job have pairwise disjoint segments. We keep them in a balanced search tree ordered by left end; a popped edge with segment $[\sigma,\tau)$ conflicts with a chosen edge at $J_a$ if and only if the chosen segment at $J_a$ with the largest left end smaller than $\tau$ ends after $\sigma$, which is tested in $O(\log n)$ time. The total running time is $O(m\log n)$, and all data structures use $O(n+m)$ space.
\end{proof}

\begin{rem}[Instance-wise guarantee]\label{rem:rho}
Let $\rho=\lambda/L\in[0,1/2]$. The proof shows $\mathit{cost}^s/\cDEG\le(1-\rho/2)/(1-\rho)$. For instance, the ratio is at most $5/4$ whenever $\rho\le1/3$; the factor $3/2$ is approached only when the relaxation pairs almost all of the processing time.
\end{rem}

\begin{coro}[Integrality gaps]\label{cor:gap}
For $g=2$, $\OPT\le\frac32\cDEG$. Hence the integrality gaps of (P), of (P) with floor cuts or blossom inequalities, of the configuration LP and of the configuration LP with time cuts are at most $3/2$. For three identical jobs, $\OPT=2$ while $\cDEG=\cCFG=3/2$, so the gap of (P) and of the configuration LP is at least $4/3$; the floor cuts and the time cuts close this example.
\end{coro}
\begin{proof}
The upper bound follows from Theorem~\ref{thm:main}, because $\cDEG$ is the smallest of these lower bounds ($\cDEG\le\cFLR\le\cMATCH$ and Lemma~\ref{lem:cfg}). For three identical unit jobs, $y_e=1/2$ on all three edges is feasible for (P) with value $3/2=L/2$, so $\cDEG=3/2$; the configuration LP has the same value, using the three pairs with $x_C=1/2$. Any schedule has one machine with two jobs and one machine with one job, so $\OPT=2$. The floor cut gives $\lambda\le1$, and the time cut requires two busy machines; both give the bound $2$.
\end{proof}

\begin{coro}[Temporal bin packing with large items]\label{cor:tbp}
Consider bins of capacity $1$ and items with sizes $q_j\in(1/3,1]$ and intervals $I_j$, where the total size of the items that a bin holds at any time may not exceed $1$ and the cost is the total time during which bins are non-empty (that is, the time horizon multiplied by the time-averaged number of open bins, the objective of~\cite{Muir24}). If $E$ is restricted to overlapping pairs with $q_a+q_b\le1$, then Lemma~\ref{lem:pair}, Lemma~\ref{lem:lp} and Theorem~\ref{thm:main} hold verbatim. In particular, LR-PAIR is a $3/2$-approximation for this problem, which contains TBP2~\cite{Muir24} and also allows items larger than half of the capacity to share bins with smaller items.
\end{coro}
\begin{proof}
Three items never fit together, so each bin holds at most two items at any time and two items that run together in a bin form an edge of the restricted set $E$; this gives part~(a) of Lemma~\ref{lem:pair}. In part~(b), (b2) shows that each component holds at most two items at any time, and (b3) shows that two items running together in a component form an edge of $F\subseteq E$, so their sizes fit. The remaining arguments use only the conflict structure of $E$.
\end{proof}

\subsection{Tightness and tie-breaking}\label{sec:tight}
Since $\tau_e=\min(c_a,c_b)$ for $e=\{J_a,J_b\}$, all edges whose segments end at the completion time of the same job have the same key; ties in the order are therefore unavoidable even when all endpoints are distinct. Unless stated otherwise, ties are broken in favour of shorter segments (larger $\sigma_e$), and all experiments in Section~\ref{sec:Results} use this rule. The guarantee of Theorem~\ref{thm:main} holds for every rule.

\begin{example}\label{ex:tight}
Consider the eight jobs $[-8,1)$ (twice), $[-4,2)$ (twice), $[-3,2)$, $[-2,2)$ and $[-1,3)$ (twice). Here $L=47$. Pairing identical jobs and pairing $[-3,2)$ with $[-2,2)$ saves $9+6+4+4=23$ and gives a schedule of cost $24$, which equals $\cDEG$, so it is optimal. LR-PAIR first processes the edges whose segments are short and end at time $1$ or $2$, and returns the four machines $\{[-8,1),[-1,3)\}$ (twice), $\{[-4,2),[-3,2)\}$ and $\{[-4,2),[-2,2)\}$, of cost $11+11+6+6=34$. The ratio is $17/12\approx1.417$.
\end{example}

The example is not an artifact of the tie-breaking rule or of coinciding endpoints. With the opposite rule (longer segments first), our local search found instances with ratio $1.4286$ ($n=10$); for instances whose endpoints are all distinct, it found an instance with ratio $1.4135$ ($n=8$), and small random perturbations of Example~\ref{ex:tight} keep the ratio above $1.33$ (Section~\ref{sec:tiesexp}; the instances are listed in Appendix~\ref{app:inst}). We do not know a family of instances on which the ratio of LR-PAIR tends to $3/2$; whether the bound $3/2$ is tight for LR-PAIR is open.

Two simple modifications keep the guarantee of Theorem~\ref{thm:main} and help in practice: running the algorithm also on the time-reversed instance ($t\mapsto-t$) and returning the better schedule, and merging two machines whenever their union is still valid, which never increases the cost because $\spn(A\cup B)\le\spn(A)+\spn(B)$. On Example~\ref{ex:tight} the two modifications together reduce the ratio to $25/24$.

\section{General Parallelism}\label{sec:general}
\subsection{A forest characterization}
A \textit{parent forest} is a set $F$ of ordered pairs $J_a\to J_b$ such that $J_a\prec J_b$, the intervals $I_a$ and $I_b$ overlap, and every job has at most one parent. An edge $J_a\to J_b$ is \textit{extending} if $c_b>c_a$ and \textit{nested} otherwise, in which case $I_b\subseteq I_a$. Its weight is $|I_a\cap I_b|$, and $w(F)$ is the total weight. For a time $t$, let $F_t$ be the set of edges of $F$ whose two jobs both run at time $t$, viewed as a graph on the vertex set $\J(t)$.

\begin{defs}\label{def:cat}
A parent forest $F$ is a \textit{$g$-caterpillar forest} if
\begin{itemize}
\item[(i)] every job has at most one extending child;
\item[(ii)] a job that is a nested child has no children;
\item[(iii)] for every time $t$, every connected component of $(\J(t),F_t)$ has at most $g$ vertices.
\end{itemize}
\end{defs}

Conditions (i) and (ii) say that each tree consists of a \textit{spine} $R_1\to R_2\to\cdots\to R_k$ of extending edges, along which start and completion times increase, and of nested leaves attached to spine jobs; this is why we call these forests caterpillars.

\begin{thm}\label{thm:cat}
For every $g\ge1$,
\[
\mathit{cost}^*_g=L-\max\{w(F): F\text{ is a }g\text{-caterpillar forest}\}.
\]
More precisely, for every $g$-caterpillar forest $F$, assigning each tree of $F$ to its own machine gives a valid schedule of cost $L-w(F)$, and every valid schedule $s$ yields a $g$-caterpillar forest $F_s$ with $w(F_s)=L-\mathit{cost}^s$. For $g=2$, conditions (i) and (ii) follow from (iii), and the theorem reduces to Lemma~\ref{lem:pair}.
\end{thm}

\begin{proof}
\emph{From schedules to forests.} Fix a machine with job set $\mathcal{A}$. For $J_b\in\mathcal{A}$, let $p(J_b)$ be, among the jobs of $\mathcal{A}$ that precede $J_b$ and overlap it, the one with the latest completion time (ties broken towards the $\prec$-smallest job); if no such job exists, $J_b$ is a root. Let $F_s$ consist of the edges $p(J_b)\to J_b$ over all machines. Each job has at most one parent and every edge goes forward in $\prec$, so $F_s$ is a parent forest. We show that it is a $g$-caterpillar forest with $w(F_s)=L-\mathit{cost}^s$: (F1) computes the weight, and (F2)--(F4) verify conditions (i), (ii) and (iii) of Definition~\ref{def:cat}.

(F1) All jobs of $\mathcal{A}$ that precede $J_b$ start no later than $s_b$, so their intersections with $I_b$ are intervals of the form $[s_b,\cdot)$, and their union is $I_b\cap I_{p(J_b)}$. Adding the jobs of $\mathcal{A}$ in the order $\prec$ and computing the length of the union shows $\spn(\mathcal{A})=\len(\mathcal{A})-\sum_{J_b}|I_b\cap I_{p(J_b)}|$, that is, $\sav(\mathcal{A})$ equals the weight of the edges of $F_s$ inside $\mathcal{A}$.

(F2) Suppose $J_a$ has two extending children $J_{c_1}\prec J_{c_2}$. Then $s_{c_2}<c_a<c_{c_1}$, so $J_{c_1}$ overlaps $J_{c_2}$, precedes it and completes later than $J_a$; hence $p(J_{c_2})\neq J_a$, a contradiction. Hence $F_s$ satisfies condition (i).

(F3) Let $p(J_b)\to J_b$ be nested and suppose $J_b$ has a child $J_c$. Then $p(J_b)\prec J_b\prec J_c$, and $p(J_b)$ overlaps $J_c$ because $s_c<c_b\le c_{p(J_b)}$. Since $c_{p(J_b)}\ge c_b$ and ties are broken towards the $\prec$-smaller job, $p(J_c)\neq J_b$, a contradiction. Hence $F_s$ satisfies condition (ii).

(F4) Let $J_b\in\mathcal{A}$ run at time $t$ and suppose that some job of $\mathcal{A}$ that precedes $J_b$ also runs at time $t$. Then $p(J_b)$ completes after $t$ and starts no later than $s_b\le t$, so it runs at time $t$ and $p(J_b)\to J_b\in(F_s)_t$. Following parents from any job of $\mathcal{A}$ running at time $t$ therefore leads, inside $(F_s)_t$, to the $\prec$-smallest job of $\mathcal{A}$ running at $t$. Hence the jobs of $\mathcal{A}$ running at time $t$ form one component of $(F_s)_t$, whose size is at most $g$. Hence $F_s$ satisfies condition (iii).

\emph{From forests to schedules.} By (i) and (ii), every tree is a spine $R_1\to\cdots\to R_k$ with nested leaves attached to spine jobs. If $R_i$ and $R_l$ ($i<l$) both run at time $t$, then so does every $R_j$ with $i<j<l$, because $s_{R_j}\le s_{R_l}\le t<c_{R_i}<c_{R_j}$; hence consecutive running spine jobs are joined by edges of $F_t$. A running leaf is contained in its spine job, which therefore also runs. Thus the jobs of a tree that run at time $t$ are connected in $F_t$, and by (iii) there are at most $g$ of them, so the machine is valid. The leaves do not change the span, and the union of the spine has length $\sum_i|R_i|-\sum_i|R_i\cap R_{i+1}|$. Hence the savings of the tree equal its weight.

\emph{The case $g=2$.} If $J_a$ had two extending children $J_{c_1}\prec J_{c_2}$, then $J_a,J_{c_1},J_{c_2}$ would all run at time $s_{c_2}$ and form a component of size three. If a nested child $J_b$ of $J_a$ had a child $J_c$, then $J_a,J_b,J_c$ would all run at time $s_c$. Thus (iii) implies (i) and (ii) for $g=2$. Moreover, for $g=2$, condition (iii) says that $F_t$ is a matching for every $t$, which is exactly compatibility.
\end{proof}

\subsection{An exact integer program and a compact relaxation}\label{sec:compact}
Theorem~\ref{thm:cat} leads to a formulation with one variable $y_{ab}\in[0,1]$ for every ordered overlapping pair $J_a\prec J_b$. We say that the pair \textit{runs during} $T\in\T$ if $T\subseteq I_a\cap I_b$, that is, if both jobs belong to $\J(T)$.
\begin{itemize}
\item[(C1)] $\sum_{a}y_{ab}\le1$ for every job $J_b$ (at most one parent);
\item[(C2)] $\sum_{b:\,J_a\to J_b\text{ extending}}y_{ab}\le1$ for every job $J_a$;
\item[(C3)] $y_{ab}+y_{bc}\le1$ for every nested pair $J_a\to J_b$ and every pair $J_b\to J_c$;
\item[(C4)] for every $T\in\T$ and every $S\subseteq\J(T)$ with $|S|=g+1$, the variables of the pairs inside $S$ sum to at most $g-1$;
\item[(C5)] for every $T\in\T$ and every $J_a\in\J(T)$, the variables of the pairs that contain $J_a$ and run during $T$ sum to at most $g-1$;
\item[(C6)] for every $T\in\T$, the variables of the pairs that run during $T$ sum to at most $N(T)-\lceil N(T)/g\rceil$.
\end{itemize}

\begin{prop}\label{prop:ip}
The $0/1$ points satisfying (C1)--(C4) are exactly the incidence vectors of $g$-caterpillar forests, and every such point satisfies (C5) and (C6). Consequently, $\max\{\sum w_{ab}y_{ab}: y\in\{0,1\}\text{ satisfies (C1)--(C6)}\}=L-\mathit{cost}^*_g$, where $w_{ab}=|I_a\cap I_b|$.
\end{prop}
\begin{proof}
Let $y$ be the incidence vector of a set $F$ of ordered overlapping pairs. Constraint (C1) says that every job has at most one parent, so $F$ is a parent forest; it is acyclic because every edge goes forward in the order $\prec$. Constraints (C2) and (C3) are conditions (i) and (ii). Fix $T\in\T$ and a time $t\in T$. The graph $F_t$ consists of the pairs of $F$ that run during $T$, and it is a forest. If a component of $F_t$ has more than $g$ vertices, it contains a subtree $S$ with exactly $g+1$ vertices and $g$ edges, which violates (C4). Conversely, if every component of $F_t$ has at most $g$ vertices, then every set $S$ of $g+1$ vertices meets at least two components, and the forest induced by $S$ has at most $|S|-2=g-1$ edges. Hence (C4) is equivalent to (iii). For (C5), the component of $J_a$ in $F_t$ has at most $g$ vertices, so $J_a$ has at most $g-1$ neighbours in it. For (C6), $F_t$ is a forest on $N(T)$ vertices whose components have at most $g$ vertices, so it has at least $\lceil N(T)/g\rceil$ components and at most $N(T)-\lceil N(T)/g\rceil$ edges. The last statement follows from Theorem~\ref{thm:cat}.
\end{proof}

Let $\cCAT$ be $L$ minus the optimal value of the linear relaxation of this program; then $\cCAT\le\mathit{cost}^*_g$. Constraint (C6) is the savings form of the time cuts. For fixed $g$ the relaxation has $O(n^2)$ variables and $O(n^{g+2})$ constraints, dominated by (C4), and it can be written down directly, without column generation. The constraints (C4) can also be generated lazily: for a given $y$ and $T\in\T$, a most violated set $S$ is a densest subgraph on $g+1$ vertices of the $y$-weighted graph on $\J(T)$, which can be found by enumeration in $O(N(T)^{g+1})$ time for fixed $g$. For $g=2$ the relaxation can be compared with those of Section~\ref{sec:lps}.

\begin{prop}\label{prop:g2}
For $g=2$, $\cFLR\le\cCAT\le\cMATCH$.
\end{prop}
\begin{proof}
Let $g=2$. Every unordered overlapping pair has exactly one orientation, so the variables of both relaxations coincide. We first show that (C1)--(C3) are implied by the constraints of (P), which coincide with (C5) for $g=2$; the argument does not extend to larger $g$, where the right-hand side of (C5) is $g-1>1$. The pairs in (C1) all contain $J_b$, and their segments $[s_b,\cdot)$ contain the elementary segment that starts at $s_b$. The extending children $J_b$ of $J_a$ overlap $J_a$ and complete after it, so their start times are smaller than $c_a$ and their segments are $[s_b,c_a)$; all these segments contain the elementary segment that starts at the latest of these start times, and all pairs contain $J_a$. In (C3), the nested child $J_b$ satisfies $I_b\subseteq I_a$ and its child $J_c$ starts in $I_b$, so both pairs contain $J_b$ and the elementary segment that starts at $s_c$. In each case the pairs share one constraint of (P). Moreover, (C5) is the constraint of (P), (C6) is the floor cut, and (C4) is the blossom inequality for sets of three jobs. Hence the feasible region of the relaxation defining $\cCAT$ contains that of $\cMATCH$ and is contained in that of $\cFLR$.
\end{proof}

For $g\ge3$ we do not know a provable relation between $\cCAT$ and $\cCFG$ or $\cCFG^+$; Section~\ref{sec:genexp} compares them experimentally.

\subsection{Why $g\ge3$ is harder}
\begin{prop}\label{prop:beta}
If an algorithm always obtains savings of at least $\beta S^*$ for some $\beta\in[0,1]$, then it is a $\bigl(1+(1-\beta)(g-1)\bigr)$-approximation.
\end{prop}
\begin{proof}
The parallelism bound gives $L\le g\cdot\OPT$, so $S^*=L-\OPT\le(g-1)\OPT$. The cost of the algorithm is at most $L-\beta S^*=\OPT+(1-\beta)S^*\le\bigl(1+(1-\beta)(g-1)\bigr)\OPT$.
\end{proof}

Thus a savings-based algorithm beats the ratio $2$ only if $\beta>(g-2)/(g-1)$, and it achieves $2-1/g$ if $\beta\ge1-1/g$. For $g=2$, $\beta=1/2$ suffices, and this is exactly what Lemma~\ref{lem:two} provides. For $g\ge3$, two obstacles remain. First, condition (iii) is not a pairwise condition. Consider the four jobs $[0,4)$, $[1,5)$, $[2,6)$ and $[3,7)$, which all run during $[3,4)$ and form a chain of three extending edges. For $g=3$, any two of the three edges can be selected together, but all three would create a component of four jobs during $[3,4)$. Hence the feasible sets are not the independent sets of a conflict graph, and the argument of Lemma~\ref{lem:two} does not apply directly. Second, even a savings guarantee of $1/2$ gives only the ratio $2$ when $g=3$.

\begin{conj}\label{conj}
For every $g\ge2$, the minimum total busy time problem with interval jobs admits a $(2-1/g)$-approximation.
\end{conj}

The conjecture is a guess based on the following observations, none of which is a proof. The ratio $2-1/g$ is achieved for proper instances by the BestCut algorithm of Mertzios \textit{et al.}~\cite{Mertzios15}; clique instances admit a polynomial-time approximation scheme~\cite{Khandekar15}; Theorem~\ref{thm:main} settles the case $g=2$; and in our experiments with $g\in\{3,4\}$, the relaxations $\cCFG^+$ and $\cCAT$ have gaps of at most $1.21$ even on instances constructed by local search (Section~\ref{sec:genexp}), far below $2-1/g$, so at least on these instances an LP-based approach is not ruled out by the integrality gap.

\section{Experimental Results}\label{sec:Results}
All algorithms and relaxations were implemented in Python~3; linear and integer programs were solved with the HiGHS solver~\cite{Huangfu18} through SciPy~\cite{Virtanen20} on a machine with two CPU cores. We use seven families of random instances, listed in Appendix~\ref{app:gen}: \textit{uniform}, \textit{short}, \textit{clique}, \textit{staircase}, \textit{bimodal}, \textit{real} and \textit{nested}. For small instances the optimum was computed by dynamic programming over subsets of jobs; for medium instances it was computed by solving the integer version of (P) with floor cuts, which is exact by Lemma~\ref{lem:pair} (Section~\ref{sec:medium} describes the tolerance and the instances that were not solved within the time limit). The code, including the random seeds and a script that checks the example of Appendix~\ref{app:mertzios}, is provided as supplementary material.

\subsection{Small instances, $g=2$}
Table~\ref{tab:small} summarizes $2{,}500$ instances with $3\le n\le11$ jobs, equally distributed over the seven families. In all instances the optimal value of the integer version of (P) coincided with the optimum (Lemma~\ref{lem:pair}), the schedule of LR-PAIR was valid, its cost equalled $L-w(F)$, and the inequalities of Theorem~\ref{thm:main} held.

\begin{table}[ht]
\caption{Small instances with $g=2$ ($2{,}500$ instances, $3\le n\le11$). All ratios are at least $1$, since the algorithms return valid schedules and $\cDEG$, $\cFLR$ and $\cMATCH$ are lower bounds on the optimum; the last column counts the instances in which the ratio is not $1$.}
\centering
\begin{tabular}{||l|c|c|c||}
\hline
Ratio & Mean & Maximum & Instances with ratio $>1$ \\ \hline
LR-PAIR / OPT & 1.0239 & 1.256 & 867 \\ \hline
FirstFit / OPT & 1.0411 & 1.500 & 1189 \\ \hline
OPT / $\cDEG$ & 1.0254 & 1.280 & 1241 \\ \hline
OPT / $\cFLR$ & 1.0000 & 1.013 & 7 \\ \hline
OPT / $\cMATCH$ & 1.0000 & 1.000 & 0 \\ \hline
\end{tabular}
\label{tab:small}
\end{table}

We also searched for bad instances by local search (random restarts followed by hill climbing on the endpoints). For LR-PAIR the largest ratios found were $1.4167$ ($n=8$, Example~\ref{ex:tight}), $1.40$ ($n=10$) and $1.406$ ($n=12$). For the relaxations, the largest ratios $\OPT/\cDEG$, $\OPT/\cFLR$ and $\OPT/\cMATCH$ found were $4/3$, $1.077$ and $1.012$, respectively. For items with sizes in $(1/3,1]$ (Corollary~\ref{cor:tbp}), $800$ further instances produced no violation of the theorem, and the largest ratio of LR-PAIR to the optimum was $1.117$.

\subsection{Tie-breaking and general position}\label{sec:tiesexp}
Table~\ref{tab:variants} compares, on $1{,}400$ instances with $4\le n\le10$, the two tie-breaking rules and the two modifications of Section~\ref{sec:tight}. Neither tie-breaking rule dominates: the rule that prefers longer segments has a slightly better mean but a worse maximum. A local search with the longer-first rule found ratios $1.40$ ($n=8$) and $1.4286$ ($n=10$), close to those of the default rule.

To test whether the bad instances depend on coinciding endpoints, we perturbed every endpoint of Example~\ref{ex:tight} independently and uniformly by at most $0.05$ ($3{,}000$ trials) and by at most $0.02$ ($6{,}000$ trials); all perturbed instances have distinct endpoints, and the largest ratios were $1.332$ and $1.333$. A local search over instances with real-valued endpoints, all distinct, found an instance with $n=8$ and ratio $1.4135$ and one with $n=10$ and ratio $1.325$. In all local searches the optimum was computed with the integer version of (P), which is exact by Lemma~\ref{lem:pair}; the instances of Appendix~\ref{app:inst} were also checked by dynamic programming over subsets.

\begin{table}[ht]
\small
\caption{Tie-breaking rules and modifications of LR-PAIR ($1{,}400$ instances, $4\le n\le10$); all of them keep the guarantee $3/2$.}
\centering
\begin{tabular}{||l|c|c|c||}
\hline
Variant & Mean / OPT & Max / OPT & Non-optimal instances \\ \hline
LR-PAIR, shorter segments first (default) & 1.0242 & 1.222 & 469 \\ \hline
LR-PAIR, longer segments first & 1.0216 & 1.242 & 455 \\ \hline
Best of LR-PAIR and its time reversal & 1.0192 & 1.222 & 422 \\ \hline
LR-PAIR with merging & 1.0187 & 1.222 & 378 \\ \hline
Both modifications & 1.0152 & 1.222 & 338 \\ \hline
\end{tabular}
\label{tab:variants}
\end{table}

\subsection{Medium instances, $g=2$}\label{sec:medium}
We generated $70$ instances with $n=100$ and $35$ instances with $n=200$, equally distributed over the seven families, and solved the integer version of (P) with floor cuts with a time limit of $120$ seconds ($n=100$) and $180$ seconds ($n=200$) per instance. We used the default relative optimality tolerance of HiGHS, $10^{-4}$; ``solved'' and ``optimum'' are meant up to this tolerance. All $70$ instances with $n=100$ were solved (mean $5.7$ seconds, maximum $38.7$ seconds). For $n=200$, $26$ of the $35$ instances were solved (mean $34.8$ seconds, maximum $144.8$ seconds). For each of the remaining nine instances, all from the uniform and nested families, the lower bound of the solver differs from $\cFLR$ by less than $0.02\%$, and the best known schedule, found by the solver or by one of the heuristics, is within $0.3\%$ of the lower bound for five instances and within $4.8\%$ to $6.0\%$ for the other four, on which the solver did not find a good schedule within the time limit. For $n=200$ we therefore divide by the best lower bound, so the reported values are upper bounds on the true ratios; they are exact for the $26$ solved instances. Restricted to these $26$ instances, the mean ratios are $1.0700$ for LR-PAIR, $1.0973$ for GreedyTracking and $1.1394$ for FirstFit. Table~\ref{tab:medium} reports the ratios, and Table~\ref{tab:family} breaks down the results by family. In addition, $\OPT/\cFLR$ was at most $1.0041$ for $n=100$ (mean $1.0004$) and at most $1.0018$ for the solved instances with $n=200$, so the floor-cut relaxation is almost exact on these instances.

\begin{table}[ht]
\caption{Medium instances with $g=2$: cost divided by the optimum ($n=100$) and by the best lower bound ($n=200$), which gives an upper bound on the ratio to the optimum.}
\centering
\begin{tabular}{||l|c|c|c|c|c||}
\hline
 & & \multicolumn{2}{c|}{$n=100$ ($70$ instances)} & \multicolumn{2}{c||}{$n=200$ ($35$ instances)} \\ \hline
Algorithm & Proven ratio & Mean & Maximum & Mean & Maximum \\ \hline
LR-PAIR & $3/2$ & \textbf{1.0745} & \textbf{1.1700} & \textbf{1.0698} & \textbf{1.1343} \\ \hline
GreedyTracking~\cite{ChangKM17} & $3$ & 1.0842 & 1.2045 & 1.0850 & 1.2149 \\ \hline
FirstFit~\cite{Flammini10} & $4$ & 1.1262 & 1.2045 & 1.1375 & 1.2149 \\ \hline
One machine per job & $2$ & 1.8008 & 1.9297 & 1.8485 & 1.9539 \\ \hline
\end{tabular}
\label{tab:medium}
\end{table}

LR-PAIR has the best mean and the smallest maximum for both sizes. Table~\ref{tab:family} shows that the advantage comes mainly from the staircase and clique families, on which LR-PAIR is best for both sizes. GreedyTracking is better on the uniform, real and nested families, by $0.9$ to $3.6$ percentage points, and the short and bimodal families are mixed. The modifications of Section~\ref{sec:tight} reduce the mean ratio of LR-PAIR for $n=100$ from $1.0745$ to $1.0712$ (better of the instance and its time reversal), $1.0676$ (merging) and $1.0648$ (both); the maximum stays at $1.1700$.

\begin{table}[ht]
\caption{Mean ratio by family ($10$ instances per family for $n=100$ and $5$ for $n=200$; denominators as in Table~\ref{tab:medium}).}
\centering
\begin{tabular}{||l|l|c|c|c|c|c|c|c||}
\hline
$n$ & Algorithm & uniform & short & clique & staircase & bimodal & real & nested \\ \hline
100 & LR-PAIR & 1.0702 & \textbf{1.0728} & \textbf{1.1345} & \textbf{1.0000} & 1.0782 & 1.0763 & 1.0897 \\ \hline
100 & GreedyTracking & \textbf{1.0568} & 1.0833 & 1.1583 & 1.0941 & \textbf{1.0722} & \textbf{1.0669} & \textbf{1.0579} \\ \hline
100 & FirstFit & 1.1302 & 1.1241 & 1.1583 & 1.1312 & 1.1336 & 1.0901 & 1.1161 \\ \hline
200 & LR-PAIR & 1.0604 & 1.0846 & \textbf{1.1192} & \textbf{1.0000} & \textbf{1.0679} & 1.0765 & 1.0798 \\ \hline
200 & GreedyTracking & \textbf{1.0515} & \textbf{1.0793} & 1.1893 & 1.0884 & 1.0855 & \textbf{1.0571} & \textbf{1.0442} \\ \hline
200 & FirstFit & 1.1479 & 1.1366 & 1.1893 & 1.1318 & 1.1479 & 1.1005 & 1.1085 \\ \hline
\end{tabular}
\label{tab:family}
\end{table}

\subsection{Running time}
LR-PAIR works on the $m=\Theta(n^2)$ edges in the worst case and is therefore slower than the greedy algorithms. Table~\ref{tab:time} reports the running times of our Python implementations on single instances: the direct implementation of Algorithm~\ref{alg:lr} and the $O(m\log n)$ implementation described in the proof of Theorem~\ref{thm:main}. The two implementations return the same schedules; we checked this on $3{,}070$ instances. The times exclude the generation of the overlap edges, which took at most $0.07$ seconds. On clique instances, doubling $n$ multiplies $nm$ by eight and the time of the direct implementation by $9.2$ to $11.2$, whereas the time of the $O(m\log n)$ implementation grows roughly like $m$. For $n=800$ the latter needs at most $0.96$ seconds, about $170$ times less than the direct implementation and about five times more than FirstFit and GreedyTracking, which need at most $0.21$ seconds.

\begin{table}[ht]
\caption{Running time in seconds of the Python implementations on single instances ($g=2$).}
\centering
\small
\begin{tabular}{||l|r|r|r|r|r|r||}
\hline
 & & & \multicolumn{2}{c|}{LR-PAIR} & & \\
Family & $n$ & $m$ & direct & $O(m\log n)$ & FirstFit & GreedyTracking \\ \hline
uniform & 100 & 2{,}291 & 0.04 & $<0.01$ & $<0.01$ & $<0.01$ \\ \hline
uniform & 200 & 9{,}571 & 0.32 & 0.01 & 0.01 & $<0.01$ \\ \hline
uniform & 400 & 40{,}731 & 3.96 & 0.13 & 0.03 & 0.02 \\ \hline
uniform & 800 & 159{,}500 & 36.09 & 0.42 & 0.12 & 0.10 \\ \hline
clique & 100 & 4{,}950 & 0.16 & 0.01 & $<0.01$ & $<0.01$ \\ \hline
clique & 200 & 19{,}900 & 1.61 & 0.05 & 0.02 & 0.02 \\ \hline
clique & 400 & 79{,}800 & 14.75 & 0.17 & 0.05 & 0.04 \\ \hline
clique & 800 & 319{,}600 & 165.55 & 0.96 & 0.21 & 0.20 \\ \hline
real & 100 & 712 & $<0.01$ & $<0.01$ & $<0.01$ & $<0.01$ \\ \hline
real & 200 & 3{,}047 & 0.04 & 0.01 & 0.01 & $<0.01$ \\ \hline
real & 400 & 11{,}658 & 0.30 & 0.02 & 0.02 & 0.01 \\ \hline
real & 800 & 47{,}005 & 2.14 & 0.11 & 0.09 & 0.06 \\ \hline
\end{tabular}
\label{tab:time}
\end{table}

\subsection{General $g$}\label{sec:genexp}
We verified Theorem~\ref{thm:cat} by brute force on $300$ instances with $6\le n\le8$ and $g\in\{2,3,4\}$: we enumerated all $1{,}197{,}738$ parent forests, of which $132{,}862$ are $g$-caterpillar forests. In every instance the maximum weight of a caterpillar forest equalled $L-\mathit{cost}^*_g$, and every caterpillar forest gave a valid schedule of cost $L-w(F)$. The integer program of Proposition~\ref{prop:ip} matched the optimum on $1{,}050$ further instances with $n\le10$ and $g\in\{2,3,4\}$.

Table~\ref{tab:lps} compares $\cCAT$ with the load bound and the configuration LPs on $600$ instances with $n\le9$. On these random instances, $\cCAT$ has much smaller gaps than $\cLOAD$ and $\cCFG$ and gaps comparable to those of $\cCFG^+$. On adversarial instances, however, $\cCFG^+$ is stronger: a local search found $\OPT/\cCAT=1.207$ for $g=3$ and $1.158$ for $g=4$ ($n=9$), whereas the same search for $\cCFG^+$ found $1.130$ and $1.125$. The advantage of $\cCAT$ is its polynomial size.

\begin{table}[ht]
\caption{Lower bounds for $g\in\{3,4\}$ ($300$ random instances each, $n\le9$): maximum of OPT divided by the bound, and number of instances with a gap. Every bound is at most OPT, so all ratios are at least $1$; an instance has a gap if its ratio is not $1$.}
\centering
\begin{tabular}{||l|c|c|c|c||}
\hline
 & \multicolumn{2}{c|}{$g=3$} & \multicolumn{2}{c||}{$g=4$} \\ \hline
Bound & Max & With gap & Max & With gap \\ \hline
$\cLOAD$ & 1.226 & 94 & 1.269 & 80 \\ \hline
$\cCFG$ & 1.188 & 108 & 1.200 & 81 \\ \hline
$\cCFG^+$ & 1.013 & 2 & 1.056 & 6 \\ \hline
$\cCAT$ & 1.042 & 8 & 1.012 & 2 \\ \hline
\end{tabular}
\label{tab:lps}
\end{table}

\section{Concluding Remarks}\label{sec:Conclusion}
We gave a combinatorial $3/2$-approximation for the minimum total busy time problem with $g=2$, which also bounds the integrality gaps of the pair relaxation and of the configuration LP by $3/2$, and we extended it to temporal bin packing with items larger than one third of the capacity. For general $g$, we characterized optimal schedules as maximum-weight $g$-caterpillar forests and derived an exact integer program and a compact relaxation. On the practical side, LR-PAIR runs in $O(m\log n)$ time, and the integer version of the pair formulation, strengthened by floor cuts, solves most instances with $200$ jobs to optimality within minutes, which makes it a useful benchmark for heuristics. Several questions remain open. For $g=2$, is $3/2$ tight for LR-PAIR, and what are the integrality gaps of $\cFLR$ and $\cMATCH$? The experiments indicate that these relaxations are much stronger than $\cDEG$; a rounding that secures two thirds of their value would give a $4/3$-approximation. For $g\ge3$, Conjecture~\ref{conj} asks for a savings guarantee of $1-1/g$; the case $g=3$, with the target ratio $5/3$, is the natural next step.

\section*{Acknowledgments}
The author used Claude (Anthropic), an AI assistant, during the preparation of this manuscript: for literature search, for drafting and editing parts of the text, for checking proofs, and for writing the code used in the experiments. The author reviewed and verified all content, including the proofs and the experimental results, and takes full responsibility for the paper.

\appendix
\section{A Remark on the Clique-Instance Bound of Mertzios \textit{et al.}}\label{app:mertzios}
Lemma~3.2 of Mertzios \textit{et al.}, in both the conference version~\cite{Mertzios12} and the journal version~\cite{Mertzios15}, states that for every fixed $g$ there is a $\frac{g\cdot H_g}{H_g+g-1}$-approximation algorithm for clique instances. The algorithm gives every set $Q\subseteq\J$ with $|Q|\le g$ the weight $\mathit{weight}(Q)=\spn(Q)-\len(Q)/g$ and runs the greedy $H_g$-approximation algorithm for weighted set cover, which repeatedly selects a set minimizing the ratio of its weight to the number of jobs it newly covers. The proof then writes the weight of the returned solution as $\sum_i\bigl(\spn(\J_i)-\len(\J_i)/g\bigr)=\mathit{cost}^s-\len(\J)/g$ and combines this with the length bound. The identity holds when the returned sets form a partition of $\J$, but the greedy algorithm returns a cover, in which a job may belong to several sets. Removing a job from a set may increase its weight, because the weight subtracts $\len(Q)/g$. The following instance shows that the stated ratio can then be exceeded.

Let $g=3$ and consider the six jobs $J_0=[-14,19)$, $J_1=[-12,4)$, $J_2=[-4,18)$, $J_3=[-13,10)$, $J_4=[-14,19)$ and $J_5=[-5,19)$, which all contain time $0$. The optimum uses the machines $\{J_0,J_4,J_5\}$ and $\{J_1,J_2,J_3\}$ and costs $33+31=64$. The greedy algorithm selects $\{J_0,J_4,J_5\}$ (weight $3$), then $\{J_0,J_3,J_4\}$ (weight $10/3$), then $\{J_0,J_2,J_4\}$ (weight $11/3$) and finally $\{J_0,J_1,J_4\}$ (weight $17/3$); in every step the minimum ratio is attained by a unique set. Every schedule obtained from this cover by assigning each job to one of the sets containing it costs at least $94$. In particular, the natural repair that assigns every job to the first selected set covering it, which corresponds to the accounting of the greedy algorithm, yields the machines $\{J_0,J_4,J_5\}$, $\{J_3\}$, $\{J_2\}$, $\{J_1\}$ with cost $33+23+22+16=94$. The ratio $94/64\approx1.469$ exceeds $\frac{3H_3}{H_3+2}=\frac{33}{23}\approx1.435$.

This observation concerns only the proof of Lemma~3.2; the other results of~\cite{Mertzios15}, such as the polynomial-time algorithm for clique instances with $g=2$ and the BestCut algorithm for proper instances, are not affected. The stated ratio itself remains achievable for clique instances, for example through the polynomial-time approximation scheme of Khandekar \textit{et al.}~\cite{Khandekar15}. The supplementary code contains a script that verifies the instance with exact rational arithmetic.

\section{Instance Generators}\label{app:gen}
Table~\ref{tab:gen} lists the generators. $U\{a,\ldots,b\}$ and $U[a,b)$ denote uniform integer and real distributions, $\mathrm{Exp}(\mu)$ the exponential distribution with mean $\mu$, and $i$ the index of the job. Small instances have integer endpoints, except for the real and nested families; medium instances have real-valued endpoints.

\begin{table}[ht]
\caption{Instance generators: start time $s$ and completion time $c$ of a job.}
\centering
\small
\begin{tabular}{||l|p{2.85in}|p{2.55in}||}
\hline
Family & Small instances ($n\le12$) & Medium instances ($n\in\{100,200\}$) \\ \hline
uniform & $s\sim U\{0,\ldots,19\}$, $c\sim U\{s+1,\ldots,20\}$ & $s\sim U[0,100)$, $c=s+U[1,60)$ \\ \hline
short & $s\sim U\{0,\ldots,18\}$, $c=s+U\{1,\ldots,5\}$ & $s\sim U[0,100)$, $c=s+U[1,12)$ \\ \hline
clique & $s=-U\{1,\ldots,20\}$, $c=U\{1,\ldots,20\}$ & $s=-U[1,50)$, $c=U[1,50)$ \\ \hline
staircase & $s=3i+U\{0,1\}$, $c=3i+6+U\{1,2,3\}$ & $s=5i+U[0,2)$, $c=5i+10+U[0,4)$ \\ \hline
bimodal & with prob.\ $0.3$: $s\sim U\{0,\ldots,5\}$, $c\sim U\{15,\ldots,20\}$; otherwise $s\sim U\{0,\ldots,18\}$, $c=s+U\{1,2,3\}$ & with prob.\ $0.25$: $c=s+U[40,90)$, otherwise $c=s+U[1,6)$; $s\sim U[0,100)$ \\ \hline
real & $s\sim U[0,20)$, $c=s+U[0,8)$ & $s\sim U[0,100)$, $c=s+\mathrm{Exp}(8)$ \\ \hline
nested & centre $z\sim U[0,20)$, half-length $h=10\,U[0,1)^2$; $s=z-h$, $c=z+h+0.01$ & centre $z\sim U[0,100)$, $h=40\,U[0,1)^2$; $s=z-h$, $c=z+h+0.01$ \\ \hline
\end{tabular}
\label{tab:gen}
\end{table}

\section{Instances from the Local Search}\label{app:inst}
The following instances are referred to in Sections~\ref{sec:tight} and~\ref{sec:tiesexp}; all of them have $g=2$.

\textbf{All endpoints distinct, default rule (ratio $1.4135$).} The eight jobs
\[
\begin{array}{llll}
{[2.5263,10.4033)}, & {[7.7087,10.3609)}, & {[5.8768,8.4170)}, & {[2.5321,10.3457)},\\
{[4.8776,5.8321)}, & {[7.5285,10.6169)}, & {[4.6105,12.9275)}, & {[8.8197,12.8037)}
\end{array}
\]
have $L=37.2269$ and $\OPT=\cDEG=19.2824$. LR-PAIR with the default rule returns a schedule of cost $27.2559$; with the opposite rule it returns cost $22.7771$ (ratio $1.181$).

\textbf{Longer segments first (ratio $10/7$).} The ten jobs $[1,13)$ (twice), $[4,6)$, $[4,9)$, $[6,8)$ (twice), $[6,9)$, $[10,11)$, $[10,12)$ and $[11,12)$ have $L=42$ and $\OPT=21$. LR-PAIR with the longer-first rule returns a schedule of cost $30$; with the default rule it returns cost $28$ (ratio $4/3$).

On both instances the other tie-breaking rule is better, so running LR-PAIR with both rules and keeping the better schedule, which keeps the guarantee of Theorem~\ref{thm:main}, helps here; we do not know whether it improves the worst case.


\begin{thebibliography}{10}
\providecommand{\url}[1]{#1}
\csname url@rmstyle\endcsname
\providecommand{\newblock}{\relax}
\providecommand{\bibinfo}[2]{#2}
\providecommand\BIBentrySTDinterwordspacing{\spaceskip=0pt\relax}
\providecommand\BIBentryALTinterwordstretchfactor{4}
\providecommand\BIBentryALTinterwordspacing{\spaceskip=\fontdimen2\font plus
\BIBentryALTinterwordstretchfactor\fontdimen3\font minus
  \fontdimen4\font\relax}
\providecommand\BIBforeignlanguage[2]{{%
\expandafter\ifx\csname l@#1\endcsname\relax
\typeout{** WARNING: IEEEtran.bst: No hyphenation pattern has been}%
\typeout{** loaded for the language `#1'. Using the pattern for}%
\typeout{** the default language instead.}%
\else
\language=\csname l@#1\endcsname
\fi
#2}}

\bibitem{AlbersH24}
S.~Albers and G.~W. van~der Heijden, ``Online busy time scheduling with
  flexible jobs,'' in \emph{Combinatorial Optimization and Applications (COCOA
  2025)}, 2025, full version: arXiv:2405.08595.


\bibitem{AlicherryBhatia03}
M.~Alicherry and R.~Bhatia, ``Line system design and a generalized coloring
  problem,'' in \emph{Algorithms -- ESA 2003}, ser. Lecture Notes in Computer
  Science, vol. 2832.\hskip 1em plus 0.5em minus 0.4em\relax Springer, 2003,
  pp. 19--30.


\bibitem{AndrewsZhang05}
M.~Andrews and L.~Zhang, ``Bounds on fiber minimization in optical networks
  with fixed fiber capacity,'' in \emph{Proceedings of IEEE INFOCOM}, 2005, pp.
  409--419.


\bibitem{Aydin20}
N.~Ayd{\i}n, {\.I}.~Muter, and {\c{S}}.~{\.I}. Birbil, ``Multi-objective
  temporal bin packing problem: An application in cloud computing,''
  \emph{Computers \& Operations Research}, vol. 121, 2020, art. no. 104959.


\bibitem{BarNoy01}
A.~Bar-Noy, R.~Bar-Yehuda, A.~Freund, J.~S. Naor, and B.~Schieber, ``A unified
  approach to approximating resource allocation and scheduling,'' \emph{Journal
  of the ACM}, vol.~48, no.~5, pp. 1069--1090, 2001.


\bibitem{Brucker07}
P.~Brucker, \emph{Scheduling Algorithms}, 5th~ed.\hskip 1em plus 0.5em minus
  0.4em\relax Springer, 2007.


\bibitem{Calinescu24}
G.~C{\u{a}}linescu, S.~Davies, S.~Khuller, and S.~Zhang, ``Online flexible busy
  time scheduling on heterogeneous machines,'' in \emph{32nd Annual European
  Symposium on Algorithms (ESA)}, ser. LIPIcs, vol. 308, 2024, art. no. 37;
  full version: arXiv:2402.11109.


\bibitem{CalinescuLiu26}
G.~C{\u{a}}linescu and M.~Liu, ``Busy time minimization with preemption,
  migration, and one resource requirement,'' arXiv:2609.23197, 2026.


\bibitem{Cao22}
N.~Cao, J.~T. Fineman, S.~Li, J.~Mestre, K.~Russell, and S.~W. Umboh, ``Nested
  active-time scheduling,'' in \emph{33rd International Symposium on Algorithms
  and Computation (ISAAC)}, ser. LIPIcs, vol. 248, 2022, pp. 36:1--36:16.


\bibitem{carli14}
T.~Carli, S.~Henriot, J.~Cohen, and J.~Tomasik, ``A packing problem approach to
  energy-aware load distribution in clouds,'' \emph{Sustainable Computing:
  Informatics and Systems}, vol.~9, pp. 20--32, 2016.


\bibitem{ChangKM17}
J.~Chang, S.~Khuller, and K.~Mukherjee, ``{LP} rounding and combinatorial
  algorithms for minimizing active and busy time,'' \emph{Journal of
  Scheduling}, vol.~20, no.~6, pp. 657--680, 2017.


\bibitem{deCauwer16}
M.~de~Cauwer, D.~Mehta, and B.~O'Sullivan, ``The temporal bin packing problem:
  An application to workload management in data centres,'' in \emph{Proceedings
  of the 28th IEEE International Conference on Tools with Artificial
  Intelligence (ICTAI)}, 2016, pp. 157--164.


\bibitem{DellAmico20}
M.~Dell'Amico, F.~Furini, and M.~Iori, ``A branch-and-price algorithm for the
  temporal bin packing problem,'' \emph{Computers \& Operations Research}, vol.
  114, 2020, art. no. 104825.


\bibitem{Edmonds65}
J.~Edmonds, ``Maximum matching and a polyhedron with 0,1-vertices,''
  \emph{Journal of Research of the National Bureau of Standards}, vol. 69B, pp.
  125--130, 1965.


\bibitem{Flammini11}
M.~Flammini, A.~Marchetti-Spaccamela, G.~Monaco, L.~Moscardelli, and S.~Zaks,
  ``On the complexity of the regenerator placement problem in optical
  networks,'' \emph{IEEE/ACM Transactions on Networking}, vol.~19, no.~2, pp.
  498--511, 2011.


\bibitem{Flammini10}
M.~Flammini, G.~Monaco, L.~Moscardelli, H.~Shachnai, M.~Shalom, T.~Tamir, and
  S.~Zaks, ``Minimizing total busy time in parallel scheduling with application
  to optical networks,'' \emph{Theoretical Computer Science}, vol. 411, no.
  40--42, pp. 3553--3562, 2010.


\bibitem{Flammini08}
M.~Flammini, G.~Monaco, L.~Moscardelli, M.~Shalom, and S.~Zaks, ``Approximating
  the traffic grooming problem with respect to {ADMs} and {OADMs},'' in
  \emph{Euro-Par 2008 -- Parallel Processing}, ser. Lecture Notes in Computer
  Science.\hskip 1em plus 0.5em minus 0.4em\relax Springer, 2008, vol. 5168,
  pp. 920--929.


\bibitem{FlamminiAlgo14}
M.~Flammini, G.~Monaco, L.~Moscardelli, M.~Shalom, and S.~Zaks, ``On the
  complexity of the regenerator cost problem in general networks with traffic
  grooming,'' \emph{Algorithmica}, vol.~68, no.~3, pp. 671--691, 2014.


\bibitem{Huangfu18}
Q.~Huangfu and J.~A.~J. Hall, ``Parallelizing the dual revised simplex
  method,'' \emph{Mathematical Programming Computation}, vol.~10, no.~1, pp.
  119--142, 2018.


\bibitem{khandekar10}
R.~Khandekar, B.~Schieber, H.~Shachnai, and T.~Tamir, ``Minimizing busy time in
  multiple machine real-time scheduling,'' in \emph{IARCS Annual Conference on
  Foundations of Software Technology and Theoretical Computer Science
  (FSTTCS)}, ser. LIPIcs, vol.~8, 2010, pp. 169--180.


\bibitem{Khandekar15}
R.~Khandekar, B.~Schieber, H.~Shachnai, and T.~Tamir, ``Real-time scheduling to
  minimize machine busy times,'' \emph{Journal of Scheduling}, vol.~18, no.~6,
  pp. 561--573, 2015.


\bibitem{KoehlerKhuller17}
F.~Koehler and S.~Khuller, ``Busy time scheduling on a bounded number of
  machines,'' in \emph{Algorithms and Data Structures (WADS)}, ser. Lecture
  Notes in Computer Science, vol. 10389.\hskip 1em plus 0.5em minus 0.4em\relax
  Springer, 2017, pp. 521--532.


\bibitem{Kolen07}
A.~W.~J. Kolen, J.~K. Lenstra, C.~H. Papadimitriou, and F.~C.~R. Spieksma,
  ``Interval scheduling: A survey,'' \emph{Naval Research Logistics}, vol.~54,
  no.~5, pp. 530--543, 2007.


\bibitem{Kovalyov07}
M.~Y. Kovalyov, C.~T. Ng, and T.~C.~E. Cheng, ``Fixed interval scheduling:
  Models, applications, computational complexity and algorithms,''
  \emph{European Journal of Operational Research}, vol. 178, no.~2, pp.
  331--342, 2007.


\bibitem{KumarRudra05}
V.~Kumar and A.~Rudra, ``Approximation algorithms for wavelength assignment,''
  in \emph{Foundations of Software Technology and Theoretical Computer Science
  (FSTTCS)}, ser. Lecture Notes in Computer Science, vol. 3821.\hskip 1em plus
  0.5em minus 0.4em\relax Springer, 2005, pp. 152--163.


\bibitem{lawler93}
E.~L. Lawler, J.~K. Lenstra, A.~H.~G. Rinnooy~Kan, and D.~B. Shmoys,
  ``Sequencing and scheduling: Algorithms and complexity,'' \emph{Handbooks in
  Operations Research and Management Science}, vol.~4, pp. 445--522, 1993.


\bibitem{Leung04}
J.~Y.-T. Leung, \emph{Handbook of Scheduling: Algorithms, Models, and
  Performance Analysis}.\hskip 1em plus 0.5em minus 0.4em\relax CRC Press,
  2004.


\bibitem{Li14}
Y.~Li, X.~Tang, and W.~Cai, ``On dynamic bin packing for resource allocation in
  the cloud,'' in \emph{Proceedings of the 26th ACM Symposium on Parallelism in
  Algorithms and Architectures (SPAA)}, 2014, pp. 2--11.


\bibitem{LiuTang22}
M.~Liu and X.~Tang, ``Busy-time scheduling on heterogeneous machines:
  Algorithms and analysis,'' \emph{IEEE Transactions on Parallel and
  Distributed Systems}, vol.~33, no.~10, pp. 3942--3958, 2022.


\bibitem{LiuTang22b}
M.~Liu and X.~Tang, ``Dynamic bin packing with predictions,'' \emph{Proceedings
  of the ACM on Measurement and Analysis of Computing Systems}, vol.~6, no.~3,
  2022, art. no. 45.


\bibitem{MartinovicS25b}
J.~Martinovic and N.~Strasdat, ``A best-cost based heuristic for busy time
  minimization,'' in \emph{Operations Research Proceedings 2023}.\hskip 1em
  plus 0.5em minus 0.4em\relax Springer, 2025, pp. 477--483.


\bibitem{MartinovicS25}
J.~Martinovic and N.~Strasdat, ``Integer linear programming formulations and
  heuristic solution approaches for busy time minimization in temporal bin
  packing,'' \emph{4OR}, vol.~24, no.~2, pp. 165--213, 2026,
  doi:10.1007/s10288-025-00602-1; preprint ``Exact and heuristic solution
  approaches for busy time minimization in temporal bin packing'',
  MATH-NM-01-2023, TU Dresden.


\bibitem{Martinovic23}
J.~Martinovic, N.~Strasdat, J.~Val{\'e}rio~de Carvalho, and F.~Furini, ``A
  combinatorial flow-based formulation for temporal bin packing problems,''
  \emph{European Journal of Operational Research}, vol. 307, no.~2, pp.
  554--574, 2023.


\bibitem{Mertzios12}
G.~B. Mertzios, M.~Shalom, A.~Voloshin, P.~W.~H. Wong, and S.~Zaks,
  ``Optimizing busy time on parallel machines,'' in \emph{Proceedings of the
  26th IEEE International Parallel and Distributed Processing Symposium
  (IPDPS)}, 2012, pp. 238--248.


\bibitem{Mertzios15}
G.~B. Mertzios, M.~Shalom, A.~Voloshin, P.~W.~H. Wong, and S.~Zaks,
  ``Optimizing busy time on parallel machines,'' \emph{Theoretical Computer
  Science}, vol. 562, pp. 524--541, 2015.


\bibitem{Muir24}
C.~Muir, L.~Marshall, and A.~Toriello, ``Temporal bin packing with
  half-capacity jobs,'' \emph{INFORMS Journal on Optimization}, vol.~6, no.~1,
  pp. 46--62, 2024, doi:10.1287/ijoo.2023.0002.


\bibitem{PadbergRao82}
M.~W. Padberg and M.~R. Rao, ``Odd minimum cut-sets and $b$-matchings,''
  \emph{Mathematics of Operations Research}, vol.~7, no.~1, pp. 67--80, 1982.


\bibitem{RenTang20}
R.~Ren and X.~Tang, ``Busy-time scheduling on heterogeneous machines,'' in
  \emph{Proceedings of the 34th IEEE International Parallel and Distributed
  Processing Symposium (IPDPS)}, 2020, pp. 306--315.


\bibitem{Saradhi10}
C.~V. Saradhi, A.~Zanardi, R.~Fedrizzi, E.~Salvadori, G.~M. Galimberti,
  A.~Tanzi, G.~Martinelli, and O.~Gerstel, ``A framework for regenerator site
  selection based on multiple paths,'' in \emph{Optical Fiber Communication
  Conference (OFC/NFOEC)}, 2010, pp. 1--3.


\bibitem{Shalom14}
M.~Shalom, A.~Voloshin, P.~W.~H. Wong, F.~C.~C. Yung, and S.~Zaks, ``Online
  optimization of busy time on parallel machines,'' \emph{Theoretical Computer
  Science}, vol. 560, pp. 190--206, 2014.


\bibitem{TianYeo15}
W.~Tian and C.~S. Yeo, ``Minimizing total busy time in offline parallel
  scheduling with application to energy efficiency in cloud computing,''
  \emph{Concurrency and Computation: Practice and Experience}, vol.~27, no.~9,
  pp. 2470--2488, 2015.


\bibitem{Vasic09}
N.~Vasi{\'c}, M.~Barisits, V.~Salzgeber, and D.~Kostic, ``Making cluster
  applications energy-aware,'' in \emph{Proceedings of the 1st Workshop on
  Automated Control for Datacenters and Clouds (ACDC)}, 2009, pp. 37--42.


\bibitem{Virtanen20}
P.~Virtanen \emph{et~al.}, ``{SciPy} 1.0: Fundamental algorithms for scientific
  computing in {Python},'' \emph{Nature Methods}, vol.~17, pp. 261--272, 2020.


\bibitem{Winkler03}
P.~Winkler and L.~Zhang, ``Wavelength assignment and generalized interval graph
  coloring,'' in \emph{Proceedings of the 14th Annual ACM-SIAM Symposium on
  Discrete Algorithms (SODA)}, 2003, pp. 830--831.


\bibitem{YeBorodin12}
Y.~Ye and A.~Borodin, ``Elimination graphs,'' \emph{ACM Transactions on
  Algorithms}, vol.~8, no.~2, 2012, art. no. 14.


\end{thebibliography}

\end{document}